%% file: main_ecrc.tex
\documentclass[5p,preprint]{elsarticle}

\graphicspath{ {./elsarticle-ecrc_JoCS/} }

\usepackage{ecrc}

\volume{00}

\firstpage{1}

\journalname{Journal of Computational Science}

\runauth{D. Bartuschat, U. R\"ude}

\jid{JoCS}

\jnltitlelogo{JoCS}

\CopyrightLine{2014}{Published by Elsevier Ltd.}

\usepackage{amssymb}

\usepackage[figuresright]{rotating}

\usepackage{mathrsfs,upgreek}

\usepackage[utf8]{inputenc}
\usepackage{amsmath}
\usepackage{amsfonts}
\usepackage{amssymb}

\usepackage[british]{babel}
\usepackage{fontenc}
\usepackage{graphicx}
\usepackage[bookmarks=false]{hyperref}
\usepackage[textwidth=\columnwidth]{todonotes}
\usepackage{booktabs}

\usepackage{subfigure}

\usepackage{multirow} 

\usepackage[ruled,vlined]{algorithm2e}
\usepackage{algorithmic}  
\usepackage{color}

\usepackage{units}
\usepackage[mode=text,per-mode=symbol,exponent-product = \cdot]{siunitx}

\usepackage{pgfplots}
\usepackage{pgfplotstable}
\usepackage{subscript}

\usetikzlibrary{backgrounds}
\pgfplotsset{every axis/.append style={
   font=\rmfamily\normalsize,
   line width=0.4pt,
   tick style={line width=0.3pt},
   legend style={font=\rmfamily\footnotesize\boldmath}
   }
}
\pgfplotsset{
   rectangLegendMarks/.style={%
      legend image code/.code={
         \draw[##1,line width=3.5pt]
            plot coordinates {
               (0pt,0pt)
               (3.5pt,0pt)
               };%
            }
      }
}

\definecolor{RYB1}{RGB}{141, 211, 199}
\definecolor{RYB2}{RGB}{255, 255, 179}
\definecolor{RYB3}{RGB}{190, 186, 218}
\definecolor{RYB4}{RGB}{251, 128, 114}
\definecolor{RYB5}{RGB}{128, 177, 211}
\definecolor{RYB6}{RGB}{253, 180, 98}
\definecolor{RYB7}{RGB}{179, 222, 105}
\definecolor{RYBred}{RGB}{225, 0,   0}
\definecolor{RYBblue}{RGB}{50, 120,   255}
\definecolor{RYBdarkblue}{RGB}{0, 0,150}
\definecolor{RYBdarkred}{RGB}{140, 0,0}

\pgfplotscreateplotcyclelist{colorbrewerWithDark_MG}{
{RYB4!95!black,fill=RYB4},
{RYB5!95!black,fill=RYB5},
{RYB2!95!black,fill=RYB2},
{RYBblue!95!black,fill=RYBblue},
{RYB3!95!black,fill=RYB3},
{violet!95!black,fill=violet},
{RYB6!95!black,fill=RYB6},
{RYBdarkred!95!black,fill=RYBdarkred},
{RYB1!95!black,fill=RYB1},
{RYBdarkblue!95!black,fill=RYBdarkblue}%
}

\pgfplotscreateplotcyclelist{colorbrewerWithDark_MG_short}{
{RYBblue!95!black,fill=RYBblue},
{RYB3!95!black,fill=RYB3},
{violet!95!black,fill=violet},
{RYB6!95!black,fill=RYB6},
{RYBdarkred!95!black,fill=RYBdarkred},
{RYB1!95!black,fill=RYB1},
{RYBdarkblue!95!black,fill=RYBdarkblue}%
}

\pgfplotscreateplotcyclelist{colorbrewerWithDark_Overall}{
{RYB3!95!black,fill=RYB3},
{RYBblue!95!black,fill=RYBblue},
{RYB1!95!black,fill=RYB1},
{RYBdarkblue!95!black,fill=RYBdarkblue},
{RYB2!95!black,fill=RYB2},
{violet!95!black,fill=violet},
{RYB6!95!black,fill=RYB6},
{RYBdarkred!95!black,fill=RYBdarkred},
{RYBred!95!black,fill=RYBred}%
}

 \newcommand{\commentGeneric}[2]{\fbox{\begin{minipage}{0.9\columnwidth}
 \textbf{Kommentar #1:}\\[1ex]
 \begin{minipage}{0.975\textwidth}
 \it #2
 \end{minipage}\end{minipage}}\quad\\[1ex]}
\newboolean{SHOWCOMMENTS}
\setboolean{SHOWCOMMENTS}{true}
\ifthenelse{\boolean{SHOWCOMMENTS}}{%
\newcommand{\commentD}[1]{\commentGeneric{Dominik}{#1}}
\newcommand{\commentU}[1]{\commentGeneric{Uli}{#1}}
}
{%
\def\commentD#1{}
\def\commentU#1{}
}

\newcommand {\lapOp} {\, \Delta \,} 
\newcommand {\gradOp} {\nabla } 

\newcommand{\eg}{\mbox{e.\,g.}}
\newcommand{\ie}{\mbox{i.\,e.}}
\newcommand{\Ie}{\mbox{I.\,e.}}
\newcommand{\cf}{\mbox{cf.}}
\newcommand{\Fig}[1]{\mbox{Fig.\,\ref{#1}}}
\newcommand{\Sect}[1]{\mbox{Sec.\,\ref{#1}}}
\newcommand{\Tab}[1]{\mbox{Tab.\,\ref{#1}}}
\newcommand{\Eqn}[1]{\mbox{Eqn.\,(\ref{#1})}}
\newcommand{\Alg}[1]{\mbox{Alg.\,\ref{#1}}}

\newcommand{\Walberla}{\textsc{waLBerla}}

\DeclareMathAlphabet{\mathpzc}{OT1}{pzc}{m}{it}
\newcommand{\pe}{$\mathpzc{pe}$}

\usepackage{listings}

\newcommand\restr[2]{{
  \left.\kern-\nulldelimiterspace 
  #1 
  \vphantom{\big|} 
  \right|_{#2} 
  }}

\begin{document}

\begin{frontmatter}



\dochead{}

\title{Parallel Multiphysics Simulations of Charged Particles in Microfluidic Flows}



\author[LSS]{Dominik Bartuschat\corref{cor1}}
\ead{dominik.bartuschat@cs.fau.de}
\cortext[cor1]{Corresponding author}
\author[LSS]{Ulrich R\"ude}
\address[LSS]{Lehrstuhl f\"ur Systemsimulation, Friedrich-Alexander Universit\"at Erlangen-N\"urnberg, Cauerstrasse 11, 91058 Erlangen, Germany}

\begin{abstract}

The article describes parallel multiphysics simulations of charged particles in microfluidic flows with the waLBerla framework. 
To this end, three physical effects are coupled: rigid body dynamics, fluid flow modelled by a lattice Boltzmann algorithm,
and electric potentials represented by a finite volume discretisation. 
For solving the finite volume discretisation for the electrostatic forces, a cell-centered multigrid algorithm is developed 
that conforms to the lattice Boltzmann meshes and the parallel communication structure of waLBerla. 
The new functionality is validated with suitable benchmark scenarios. 
Additionally, the parallel scaling and the numerical efficiency of the algorithms are analysed on an advanced supercomputer.
\end{abstract}


\end{frontmatter}


\input{intro}
\input{numerical_modeling}
\input{mg_solver}
\input{walberla_bc}
\input{validation}
\input{results}
\input{conclusion}
\section*{Acknowledgements}
The authors would like to thank 
Simon Bogner, Tobias Preclik, Daniel Ritter, Ehsan Fattahi, and Bj\"orn Gmeiner for valuable discussions,
and Christian Kuschel, Markus Huber, and Gaby Fleig for support in the correction process.
The authors are grateful to the RRZE and LRZ for providing the computational resources on LiMa and SuperMUC, respectively.






\section*{References}
\bibliographystyle{elsarticle-num}
\bibliography{bibfile_June2013}







\end{document}

%% file: intro.tex
\section{Introduction\label{sec:Introduction}}
Computer simulations incorporating and coupling
multiple physical effects rapidly gain importance in science and engineering.
They can be used to predict and optimise the behaviour of processes or devices
in engineering applications.
However, the high computational complexity of such multiphysics simulations often requires
the use of parallel supercomputers when realistic scenarios are studied.

In this paper, we develop a simulation model for
charged objects in fluids that are additionally subject to electric fields.
Systems of this type occur in a wide range of applications.
For example, electrostatic filters can be designed, such that
particles are removed from non-conductive liquids such
as oil~\cite{yanada2008fundamental}
or for filtering pollutants from exhaust gases.
Medical scenarios may involve charged particles in aerosols for pulmonary drug delivery~\cite{Bailey19983,Longest2012_296} and in
lab-on-a-chip systems where charged objects, such as cells or DNA, can be manipulated and separated by means of
electric fields~\cite{KangLi:2009:EKFParticlesCells}.

\begin{figure}[htb]
  \centering
     \includegraphics[width=0.48\textwidth,bb=14 14 1295 975,trim={0 0 0 6.2cm},clip]{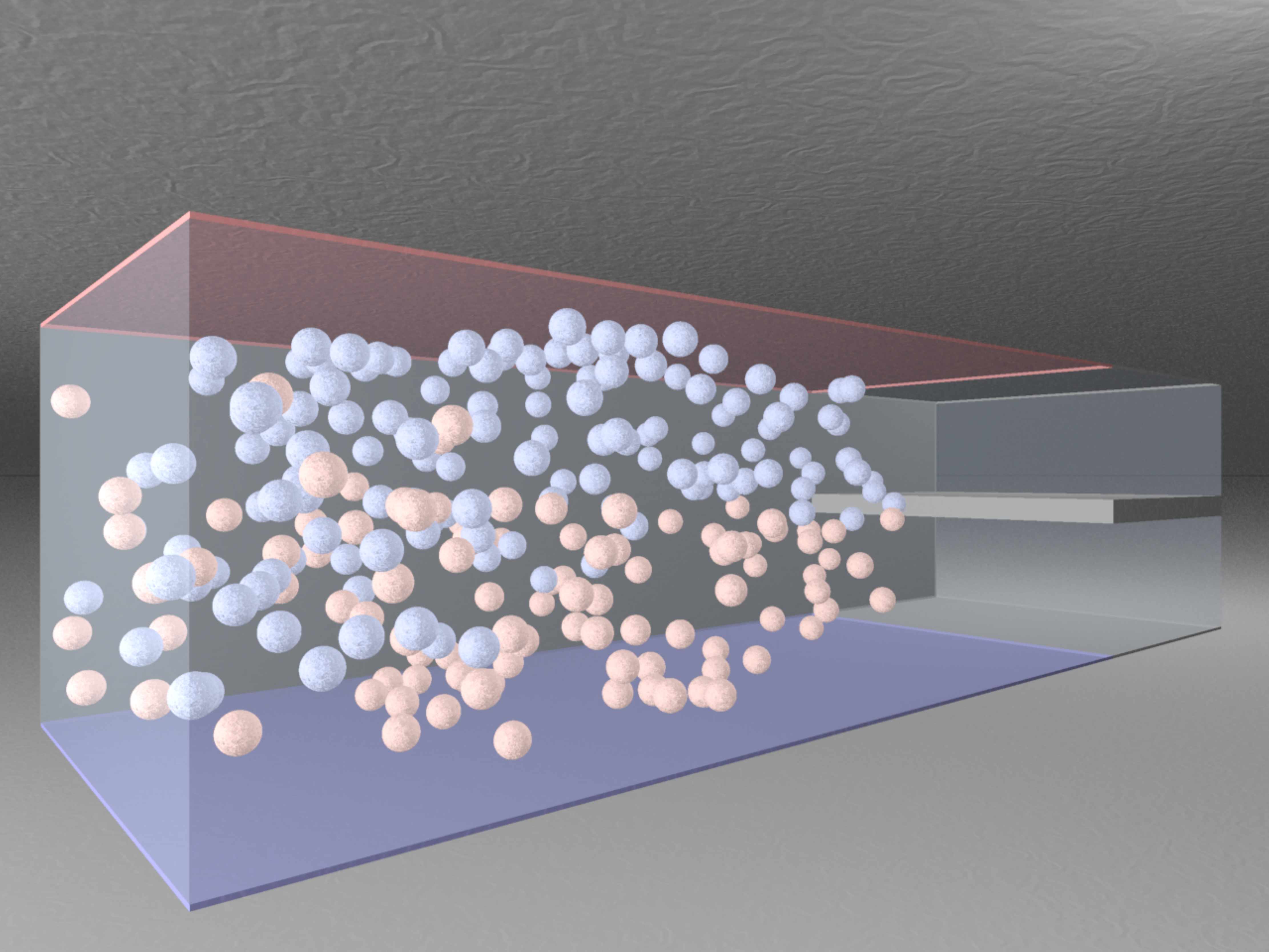}
    \caption{Separation of oppositely charged particles in a micro-channel fluid flow.}
  \label{fig:CP_Separ_250Intro}
\end{figure}

For simulating these scenarios, it is necessary to model the coupling between
three system components:
fluid flow, charged objects, and electric fields.
This paper presents the design and implementation of efficient parallel algorithms
for the simulation of such a coupled multiphysics constellation on advanced high performance computers.
As example setup, we model a micro-channel flow of homogeneously charged
rigid particles moving in a fluid, subject to an electrostatic field 
that is applied perpendicular to the flow direction.
In this situation, the charged particles are
transported with the fluid and are deflected by the electrostatic forces.
We simulate two cases:
the agglomeration of equally charged particles on an oppositely charged channel wall,
and the separation of oppositely charged particles in a bifurcating micro-channel, as depicted in~\Fig{fig:CP_Separ_250Intro}.
Double-layer effects will not be considered in this paper, \ie{}, we treat the H\"uckel limit of electrophoresis.

\begin{figure}[htb]
  \centering
     \includegraphics[width=0.48\textwidth,bb=14 14 1017 565]{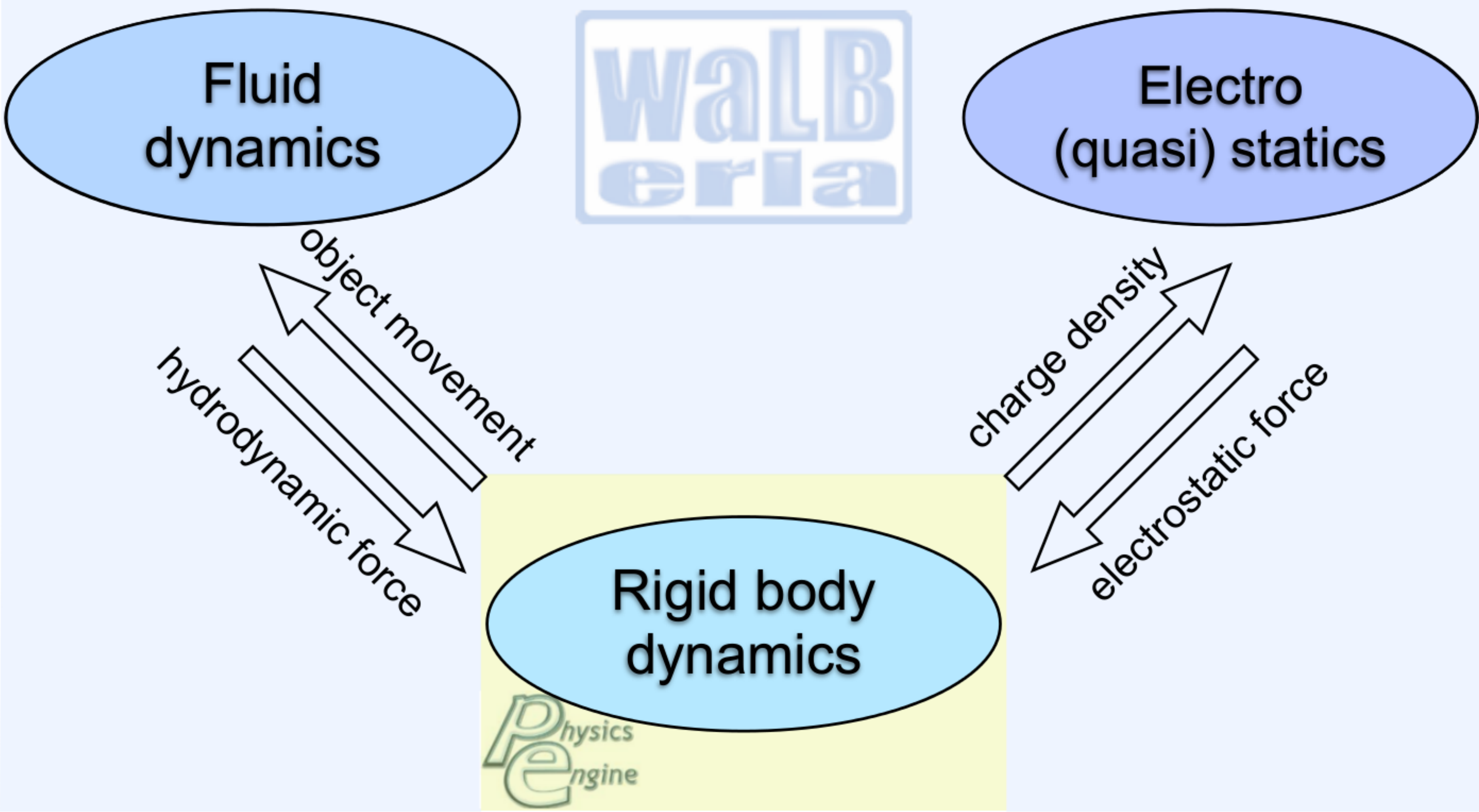}  
    \caption{Coupled physical effects simulated with~\Walberla{} and the~\pe{}.\label{fig:CP_Multiphysics_Interactions}}
\end{figure}

The simulation scenario and the physical models are realised within~\Walberla{}~\cite{Donath:2008:KONWIHR07,Feichtinger2011105} (see \Fig{fig:CP_Multiphysics_Interactions}).
\Walberla{} is a framework for massively parallel simulations of fluid flow applications that employ the lattice Boltzmann method (LBM)~\cite{Chen:98:LBM_For_Fluid_Flows,aidun2010lattice}.
The LBM can be used as an alternative to classical continuum mechanics approaches that has special advantages when simulating complex flow phenomena, such as moving objects.
LBM is structurally an explicit time stepping method and has local data dependencies. 
Thus, it can be parallelised with high efficiency even on large processor numbers.
\Walberla{}'s modular design permits the flexible extension with internal modules and the
coupling with external libraries.
For simulations involving the interaction of fluids and rigid bodies~\cite{Goetz:2010:ParComp},
\Walberla{} can be coupled with the \pe{} physics engine~\cite{Goetz:2010:SC10,Iglberger:2009:CSRD}.
The particles are treated as fully resolved rigid geometric objects.
We use the \pe{}'s parallel fast frictional dynamics algorithm~\cite{iglberger:2010:Pe}
for simulating rigid body dynamics.

In the LBM, we use the two-relaxation-time (TRT) model of Ginzburg~et~al.~\cite{ginzburg2008two}.
The LBM and the rigid body simulation are coupled with a four-way interaction, as developed in G\"otz~et~al.~\cite{Goetz:2010:ParComp}.
This uses the momentum exchange method of Yu~et~al.~\cite{2003:LBM:Yu} similar to Ladd~\cite{Ladd_1993_PartSuspPt1} and Aidun~et~al.~\cite{Aidun:98:ParticulateSuspensions}.
Since the LBM per se can represent the forces between particles 
only correctly if the separation exceeds one lattice spacing~\cite{Ladd_1994_PartSuspPt2},
we employ a {\em lubrication correction} as introduced in Nguyen and Ladd~\cite{nguyen2002lubrication}.

In this paper, we describe how~\Walberla{} is augmented to additionally solve
elliptic partial differential equations
as they can be used
to represent
electric potentials,
and how this is coupled to the LBM and the \pe{} for particle transport.
This requires the design and implementation of additional new functionality
for boundary condition handling and efficiently solving large sparse linear systems of equations.

For discretising the electric potential equation,
we choose a finite volume discretisation whose mesh spacing is conforming to the LBM grid.
For the solution of the sparse linear systems, we develop
a fast geometric multigrid (MG) algorithm.
Fitting to \Walberla{}'s parallelisation and data layout, a
cell-centered MG method~\cite{Mohr:2004:CCMR} is used.
Galerkin coarsening is applied to provide a robust solver that is easy to use for different governing equations and boundary conditions (BCs).

In our work, achieving excellent computational performance is a primary goal.
We gain additional speedups by using a representation of the linear
system based on a {\em stencil} paradigm and using systematically that these stencils are almost uniformly identical throughout the simulation domain.
Special stencils need to be used only near boundaries or material interfaces.
For a discussion of various parallel implementation and performance optimisation approaches for MG methods, we refer to~\cite{2011:Gmeiner,Huelsemann:2005:PGM}.

The physical correctness of our simulations is validated for each of the coupled models.
This includes fluid-particle interaction at low Reynolds numbers with lubrication correction and the electrostatic forces on the particles.
Parameter studies show the impact of model parameters on physical accuracy.

Finally, we present performance results on the SuperMUC cluster of the Leibnitz Supercomputing
Centre\footnotemark[1]\footnotetext[1]{\url{http://www.lrz.de/services/compute/supermuc/}} (LRZ) in Garching, Germany.
Good scaling behaviour is shown for more than $68 \cdot 10^{9}$ lattice cells on up to 32,768 cores. 
In this case, more than seven million fully resolved charged objects that constitute for 9.4\% solid volume fraction 
interact with the flow and exhibit electrostatic effects. 

%% file: numerical_modeling.tex
\section{Numerical Modelling}
\label{Sec:NumModeling}
\subsection{Lattice Boltzmann Method}
\label{SubSec:TRT}
The LBM is based on kinetic theory
for gases, describing molecular motion.
However, in contrast to molecular dynamics, where individual molecules are considered,
the LBM statistically describes ensembles of molecules.
Particle\footnotemark[2]\footnotetext[2]{These virtual fluid particles in the LBM context are different from the simulated macroscopic particles}
distribution functions (PDFs) represent the spatial and velocity
distribution of molecules in \textit{phase space} over time.
The dynamics is described by the Boltzmann transport equation in terms of the
temporal and spatial variation of PDFs, balanced by molecular collisions.
The solution of the Boltzmann equation converges towards the Maxwell-Boltzmann distribution of molecules that are in local thermodynamic equilibrium.
At any instant of time, macroscopic properties can be computed as moments of the PDFs~\cite{succi2001lattice,hanel2004molekulare}.

For the LBM,
the phase space is discretised into a Cartesian lattice $\Omega_{\delta x} \subset \mathbb{R}^D$ of dimension $D$ with spacing $\delta x$,
and a finite set of $Q$ discrete velocities $\vec{c}_q \in \mathbb{R}^D, q \in \{1,\ldots,Q\}$.
These are chosen such that within a time increment $\delta t = t_{n+1} - t_n$
with discrete time $T_{\delta t} = \{ t_n: n = 0,1,2,\ldots \} \subset \mathbb{R}^+_0$,
molecules can move to adjacent lattice sites or stay.
Associated with each $\vec{c}_q$ is a PDF $f_q: \Omega_{\delta x} \times T_{\delta t} \mapsto \mathbb{R}$.
Explicit discretisation in time and finite difference discretisation in space~\cite{wolf2000lattice}
result in the discrete lattice Boltzmann equation
\begin{equation}
  f_q(\vec{x}_i + \vec{c}_q \delta t, t_n + \delta t) - f_q(\vec{x}_i,t_n) = \delta t \mathcal{C}_q + \delta t F_q.
  \label{Eq:discrLBE_extForce}
\end{equation}
This equation describes the streaming of the PDFs between neighbouring lattice sites
and the subsequent collisions, represented by the collision operator $\mathcal{C}_q$.
The external body-force term $F_q$ will be disregarded in this section.

The LBM can be employed with different collision operators.
The simplest collision operator is the BGK model~\cite{1954:BGKPhysRev.94.511}
that linearly relaxes the PDFs towards equilibrium in velocity space, with a single dimensionless relaxation time $\tau$ (or collision frequency $\omega=\tau^{-1}$).
However, the BGK has disadvantages regarding its stability and accuracy~\cite{Luo2011CollModEff}.
In particular, the wall locations for the standard implementation of no-slip BCs depend on $\tau$:
These BCs of first order enforce zero velocity at the wall by setting
\begin{equation}
   f_{\bar{q}}( \vec{x}_f, t_n + \delta t) = \tilde{f}_{q}(\vec{x}_f,t_n).
   \label{Eq:BBBC}
\end{equation}
Consequently, PDFs of a fluid cell $\vec{x}_f$ adjacent to a wall are reflected from the wall to direction $\bar{q}$ opposite to their original direction $q$.

A more sophisticated operator with multiple relaxation times was presented in d'Humi\`{e}res~\cite{dHum1992}.
It performs collisions in moment space, relaxing the different moments towards their equilibrium.
This method is numerically stable and accurate~\cite{Luo2011CollModEff},
but computationally costly.

We use the stable, accurate, and computationally cheaper
TRT collision operator~\cite{ginzburg2004lattice,ginzburg2008two} for collisions in velocity space
\begin{equation}
  \mathcal{C}_q =
    \lambda_e \left( f^e_q - f_q^{\text{eq},e} \right) +
    \lambda_o \left( f^o_q - f_q^{\text{eq},o} \right),
  \label{Eq:TRTOp}
\end{equation}
with two relaxation times, $\lambda_e$ for even- and $\lambda_o$ for odd-order non-conserved moments.
The parameters $\lambda_e$ and $\lambda_o$ are eigenvalues of the TRT collision matrix $\mathbf{S}$
that is related to the collision operator by $\vec{\mathcal{C}} = \mathbf{S} \left( \vec{f} - \vec{f}^\text{eq} \right)$,
with $\vec{f} := \left(f_0, \ldots f_{Q-1} \right)^T \in \mathbb{R}^Q$ and $\vec{f}^\text{eq}$, $\vec{\mathcal{C}}$ defined analogeously.
$\lambda_e=-\tau^{-1}$ 
is related to the kinematic viscosity $\nu = \left(\tau - \frac{1}{2}\right) c_{s}^2 \delta t$,
whereas $\lambda_o$ can be chosen freely.
To ensure stability, $\lambda_o$ should be within the interval ${]-2,0[}$.
Ginz\-burg~et~al. \cite{ginzburg2008study} use the parameter $\Lambda = \left( \frac{1}{2} + \frac{1}{\lambda_e} \right)\left( \frac{1}{2} + \frac{1}{\lambda_o} \right)$:
For bounce-back BCs (\Eqn{Eq:BBBC}), walls are fixed midway between two lattice sites, if $\Lambda:=\Lambda_\text{mid} = \frac{3}{16}$, independent of $\tau$.
For porous media, the optimal value is $\Lambda:=\Lambda_\text{por} = \frac{1}{4}$.

TRT requires the decomposition $f_q = f^e_q + f^o_q$ into even and odd components
\begin{equation}
  \begin{array}{l c r}
  f^e_q = \frac{1}{2} ( f_q + f_{\bar{q}} ) &\text{ and }& f^{\text{eq},e}_q = \frac{1}{2} ( f^\text{eq}_q + f^\text{eq}_{\bar{q}} ) \\
  f^o_q = \frac{1}{2} ( f_q - f_{\bar{q}} ) &\text{ and }& f^{\text{eq},o}_q = \frac{1}{2} ( f^\text{eq}_q - f^\text{eq}_{\bar{q}} ),
  \end{array}
\end{equation}
with $\vec{c}_q = - \vec{c}_{\bar{q}}$.
The local equilibrium distribution function for incompressible LBM, according to He and Luo~\cite{HeLuo:97}, is then given for each lattice site by
\begin{equation}
  \begin{array}{l c}      
      f^{\text{eq},e}_q = w_q \left( \rho_f - \frac{\rho_0}{2c_{s}^2}({\vec{u}} \cdot \vec{u}) + \frac{\rho_0}{2c_{s}^4}({\vec{c}_q} \cdot \vec{u})^2 \right) \\
      f^{\text{eq},o}_q = w_q \frac{\rho_0}{c_{s}^2}(\vec{c}_q \cdot \vec{u}),
  \end{array}
\end{equation}
where `$\cdot$' denotes the standard Euclidean scalar product.
Both, macroscopic density $\rho_f = \rho_0 + \delta\rho$ with fluctuation $\delta\rho$, and velocity $\vec{u}$, 
can then be computed from moments of $f$
\begin{equation}
  \begin{array}{l c}
      \rho_f (\vec{x}_i, t) = \sum\limits_{q} f_q(\vec{x}_i, t), \\
      \vec{u}(\vec{x}_i, t) = \frac{1}{\rho_0} \sum\limits_{q} \vec{c}_{q} f_q(\vec{x}_i, t).
  \end{array}
\end{equation}
We use the D3Q19 model of Qian~et~al.~\cite{1992:QianBGK}, where the thermodynamic speed of sound is $c_{s}= c/\sqrt{3}$, with ${c = \delta x / \delta t}$.
Then, the weights $w_q$ are: ${w_1 = 1/3}$, ${w_{2, \ldots, 7} = 1/18}$, and $w_{8,\ldots, 19} = 1/36$.

In each time step $t_n \in T_h$ the lattice Boltzmann method performs a \emph{collide--} and a \emph{stream step}
\begin{equation}
  \hspace{-0.2cm}
  \begin{array}{r@{\hspace{0.5ex}}l}
  \tilde{f}_q(\vec{x}_i,t_n) = f_q(\vec{x}_i,t_n) &+ \lambda_e [ f^e_q(\vec{x}_i,t_n) - f_q^{\text{eq},e}(\vec{x}_i,t_n) ] \\ 
                                                  &+ \lambda_o [ f^o_q(\vec{x}_i,t_n) - f_q^{\text{eq},o}(\vec{x}_i,t_n) ]
 \end{array}
\end{equation}
\begin{equation}
f_q(\vec{x}_i + \vec{e}_q, t_n+\delta t) = \tilde{f}_q(\vec{x}_i,t_n),
\end{equation}
where $\tilde{f}_q$ denotes the post-collision state and $\vec{e}_q =\vec{c}_q \delta t$.
In the collide step,
the PDFs are relaxed towards equilibrium, whereas the stream step models mass transport.

Computationally, the collide step is compute-intensive
while the stream step is memory-intensive.
To increase efficiency, both steps are fused to a \emph{stream-collide} step.
Since PDFs are only streamed between adjacent lattice sites, LBM has a strictly local memory access pattern.
Thus, it is parallelisable, and carefully designed implementations scale
well on parallel architectures.

To ensure consistent units in the computations, \Walberla{} converts physical entities to lattice units (LUs) at the beginning of the simulation.
In the following, physical parameters in LUs are indicated with subscript $L$, \eg{} $\delta x_L$.
The parameters $\delta x_L$, $\delta t_L$, $\rho_{f,L}$ and the electric potential $\Phi_L$ are thus normalised to unity.

\subsection{Fluid-Particle Interaction} \label{SubSec:FPI}
Particles suspended in fluids get dragged along with a flow due to momentum transferred from the fluid.
They in turn influence the flow by transferring momentum to the fluid.
Methods based on the Navier-Stokes equation usually model the interaction by imposing the continuity of normal stresses and velocities at fluid-object interfaces.
Since the LBM is a mesoscopic method, momentum transferred from the fluid to objects can be directly computed from PDFs.
The continuity of velocities can be imposed by no-slip BCs on the surface of the objects.
This interaction is modelled in~\Walberla{} by the momentum exchange approach
that was first presented in Ladd~\cite{Ladd_1993_PartSuspPt1,Ladd_1994_PartSuspPt2}, with particles treated as fluid-filled shells.
Due to reduced stability, and because this method is limited to particles whose density is larger than the fluid density,
Aidun et~al.~\cite{Aidun:98:ParticulateSuspensions} suggest to treat the particles as solid objects.

Our method follows Yu~et~al.~\cite{2003:LBM:Yu} and considers the particles as solid, rigid objects that are mapped to the lattice.
Each cell whose center is overlapped by a particle is regarded as solid moving obstacle cell.
The other cells are fluid cells, on which the LBM is applied with BCs modelling the moving objects.
This method is efficient and easy to implement because the force acting on an object can be computed directly from PDFs in the surrounding cells.
The solid particles come at the cost of fluid cells appearing and disappearing due to particle movement.
Consequently, the PDFs of fluid cells must be reconstructed:
we set the PDFs of the wake-cell to the equilibrium distribution based on the object surface velocity and the lattice fluid density $\rho_{f,L} = 1$.

Particles transfer momentum to the fluid by means of accelerating no-slip BCs introduced in Ladd~\cite{Ladd_1993_PartSuspPt1}.
The fluid velocity at a given fluid cell $i$ at $\vec{x}_{i}$ adjacent to the particle
is matched to the local velocity $\vec{u}_{s}$ at a particle surface cell $s$ at $\vec{x}_{s}$ by setting
\begin{equation}
   f_{\bar{q}} \left( \vec{x}_{i} , t_n \right) = \tilde{f}_q \left( \vec{x}_{i} , t_n \right) - 2 \frac{\omega_q}{c_s^2} \rho_{f} \vec{c}_{q} \cdot \vec{u}_{s}.
   \label{Eq:LBMMovWall}
\end{equation}
Here, the PDF reflected from the particle surface opposite to the direction $q$ pointing from $\vec{x}_{i}$ to $\vec{x}_{s}$ is modified accordingly.
$\vec{u}_{s}$ describes the object movement, including rotation and translation.
Representing object boundaries by a {\em staircase approximation} is only
first-order accurate,
but it is computationally efficient and is sufficient for our applications as will be shown in \Sect{sec:Validation}.
Higher-order BCs are either computationally more demanding, such as multi-reflection in Ginzburg~et~al.~\cite{2003:Ginzburg:PhysRevE.68.066614},
or yield $\tau$-dependent results, such as Bouzidi BCs~\cite{bouzidi2001momentum}.
More details on this viscosity dependence can be found in~\cite{d'Humieres:2009:VIN:1576854.1576919,bogner2014drag}.

The momentum transferred from fluid to the object can be computed
from the difference in momentum density before and after the reflection of PDFs at the solid object.
Summing up the momenta transferred from fluid cells $i$ to
neighbouring particle surface cells $s$,
the hydrodynamic force on the particle can be computed following~\cite{mei2002force} as
\begin{equation}
   \vec{F}_{h} = \sum\limits_{s} \sum\limits_{q \in D_s } \vec{c}_q \left[ 2 \tilde{f}_q \left( \vec{x}_{i}, t_n \right) - 2 \frac{\omega_q}{c_s^2} \rho_{f} \vec{c}_{q} \cdot \vec{u}_{s} \right] \frac{\delta x^3}{\delta t}.
   \label{Eq:LBMhydrodynForce}
\end{equation}
Here, $\vec{x}_i = \vec{x}_s + \vec{e}_{\bar{q}}$ and the set $D_s$ of direction indices $q$, in which a given $s$ is accessed from adjacent $i$.

The hydrodynamic force is summed up at the boundary sites of each solid object.
These partial forces are then aggregated to the total force acting on the center of mass and the corresponding torque.
In the subsequent position update for the objects, both the translation and rotation are computed.
They in turn affect the fluid via $\vec{u}_{s}$ and that way other particles.
This interaction of fluid and particles alone corresponds to a two-way coupling.
Since additionally the interaction of individual particles and of particles and walls 
is fully modelled in the \pe{} algorithms, we achieve a full four-way coupling.

\subsection{Lubrication Correction} \label{SubSec:LubrTheory}
Lubrication forces, defined here following Bossis and Brady~\cite{bossis:1984:DynSimShrSusp_jcp} and Cox~\cite{cox:1974:motionSuspPart},
are strong forces occuring when particles in a fluid move relative to each other in close proximity.
In case of approaching particles, fluid in the gap between the particles must be displaced, and a high pressure arises.
The corresponding force is increasing with larger relative velocity and decreasing gap width.
Lubrication forces counteract the particle movement, \ie{}, they are attractive when particles move apart.

An insufficient handling of these lubrication forces can result in wrong particle interactions, such as particles unphysically sticking together in the simulation.
The LBM with moving objects can accurately capture lubrication forces for particle separation distances larger than one lattice site for creeping flow~\cite{Ladd_1994_PartSuspPt2}.
For smaller gap sizes, the increasing force can not be recovered directly. Instead, the force stays approximately constant.

As shown in Ladd~\cite{Ladd_1994_PartSuspPt2} for spheres of equal size,
the lubrication force depends on $s = \frac{h_{g}}{R}$, the gap width normalised by the sphere radius.
The normal lubrication force acting between the particle centers is proportional to $s^{-1}$.
Tangential lubrication forces are proportional to $\ln(s^{-1})$.
Due to the logarithmic dependency, their effect is much weaker for small gaps so that they are neglected in the present paper.
We refer to results of Janoschek~et~al.~\cite{janoschek2013accurate}, which show that for solid volume fractions below 30\%,
tangential lubrication correction can be neglected to accurately simulate shear-induced diffusion and suspension viscosity with spherical particles.

The lubrication correction for the LBM was introduced in Ladd~\cite{Ladd:1997:Sedimentation} for the normal force between two equally sized spheres:
By correcting the computed force on the spheres with the part of the lubrication force not captured by LBM, for distances below a threshold $h_c$ in the order of one lattice spacing.
It adds the difference between the analytical solution of the lubrication force~\cite{claeys:1989:lubrication,cox:1974:motionSuspPart} in Stokes flow
and the lubrication force at the threshold distance that is still captured by LBM,
replacing $\frac{1}{s}$ by $(\frac{1}{s} - \frac{R}{h_c})$.

An extension for two differently sized spheres was presented in Ladd and Verberg~\cite{Ladd:2001:LBsimSuspLubr}
\begin{equation}
  \vec{F}^{Lub}_{a\,b} =
\left\{
  \begin{array}{l l}
    6 \pi \eta  \frac{ R_a^2 \cdot R_b^2 }{ (R_a + R_b)^2 } \left( \frac{1}{h_{g}} - \frac{1}{h_{c}} \right)  u^{rel}_{n} \hat{r}_{a\,b} & \text{, if } h_{g} \leq h_{c} \\
    0 & \text{, else} \\
  \end{array} \right.
\label{Eq:NormalLubrCorr}
\end{equation}
with the relative particle velocity in direction of the contact normal $u^{rel}_{n} =  \vec{u}_{a\,b} \cdot \hat{r}_{a\,b}$ 
given by the inner product of the velocity difference $\vec{u}_{a\,b} = \vec{u}_b - \vec{u}_a$ of both spheres
and the normalised distance vector $\hat{r}_{a\,b} = \frac{\vec{r}_{a\,b}}{ \lvert \vec{r}_{a\,b} \lvert }$.
Here, $\vec{r}_{a\,b} = \vec{x}_b - \vec{x}_a$ is the vector connecting the centers of both spheres with radius $R_a$ and $R_b$.
For gaps $h_{g}$ larger than the cut-off distance $h_{c}$, the lubrication correction force is not applied.
Sphere-wall lubrication correction can be computed with this formula by letting \eg{} ${R_b \rightarrow \infty}$ and for stationary walls setting $\vec{u}_b = 0$.

We perform normal lubrication force correction by adding the force in~\Eqn{Eq:NormalLubrCorr} due to surrounding particles to the computed hydrodynamic force of each particle.
Following Nguyen and Ladd~\cite{nguyen2002lubrication}, the lubrication correction term is added for each pair of particles.
Nguyen and Ladd refer to Brady and Bossis~\cite{brady1988stokesian}, who demonstrate that this yields sufficiently accurate results for their method.

To ensure stability of the simulation, the following restrictions were added:
The lubrication force is limited to a maximum value, preventing the system from getting too stiff, which would significantly restrict the time-step size.
For this purpose, the gap width in the lubrication force computation is restricted to a minimum threshold value ($0.01 \cdot \delta x$ in all following simulations).
This minimum gap is by a factor of $R$ 
smaller than the distance where qualitatively important physics typically still occurs -- at gaps down to $0.01R$~\cite{nguyen2002lubrication}.
Another limiter was introduced for particles with a high separation velocity that can occur after the collision of particles.
In case $u^{rel}_{n} > u^{sep}_{max}$, the normal lubrication force is limited by rescaling it's magnitude ${F}^{Lub}$ as
$\vec{F}^{Lub} = \frac{\vec{F}^{Lub}}{ {F}^{Lub} } \left( 1 + \log_{10}( {F}^{Lub} ) \right)$.
This limiter prevents problems that occur with the parallel fast frictional dynamics algorithm~\cite{iglberger:2010:Pe} as it is used in the \pe{} and which
resolves collisions by computing a post-collision velocity to push the particles apart.
The lubrication force computed in the next time-step acts opposite to that movement, due to the tiny gap and high separation velocity.
This leads to particles being pushed apart strongly, possibly resulting in stability problems.

All methods above use only the dominating singluar terms of lubrication force components.
Recently, an even more elaborate method for lubrication correction of a\-spherical particles was presented in~\cite{janoschek2013accurate}, 
taking into account higher-order singular terms.
However, this is not incorporated in the present algorithm.

\subsection{Electric Potential}
\label{SubSec:ElPot}
The interaction of charged particles and of these particles with charged walls can be modelled by the electric potential
and resulting Coulomb forces acting on the objects.
The spatially varying electric potential $\Phi(\vec{x})$ produced by the charge density $\rho(\vec{x})$
is described by Poisson's equation
\begin{equation}
   - \lapOp \Phi (\vec{x}) = \frac{\rho(\vec{x})}{\varepsilon}
   \label{Eq:Poisson}
\end{equation}
for spatially constant permittivity $\varepsilon$.
The charge density is zero in the fluid and nonzero at the particle locations.
\paragraph{Finite volume discretisation}
In order to solve \Eqn{Eq:Poisson}, we apply a conservative finite volume~\cite{hirsch2007numerical,Eymard2000FVM} scheme
on the regular LBM lattice that subdivides the spatial domain into cells which act as control volumes.
This includes volume integration over each cell and applying the divergence theorem.
The resulting sum over the fluxes $\gradOp \Phi = -\vec{E}$ across the cell surfaces then
balances the volume integral over the right-hand side (RHS) of \Eqn{Eq:Poisson}, conserving charge.
When the fluxes are approximated by central differences of $\Phi_{i}$ from the neighbouring and current cell,
one obtains the seven-point stencil for each unknown $\Phi_{i}$
\begin{equation}
\hspace{-4pt}
 \setlength{\arraycolsep}{1.2pt}
  \lapOp_{\delta x} = 
  \frac{-1}{{\delta x}^2}
  \left[
   \begin{matrix}
      \begin{pmatrix} 
        0 & ~0 & ~0 \\
        0 & -1 & ~0 \\
        0 & ~0 & ~0      
      \end{pmatrix}
      \begin{pmatrix} 
        ~0 & -1 & ~0 \\
        -1 & ~6 & -1 \\
        ~0 & -1 & ~0
      \end{pmatrix}
      \begin{pmatrix} 
        0 & ~0 & ~0 \\
        0 & -1 & ~0 \\
        0 & ~0 & ~0      
      \end{pmatrix}
   \end{matrix}
  \right]
\end{equation}
for uniform lattice spacing $\delta x$. 
This scheme provides up to $\mathcal{O}({\delta x}^2)$ accuracy.
The unknowns are associated with the cell centers and represent the mean value over the cell.
Consequences of this approach to the solver for the resulting linear system of equations are discussed in \Sect{sec:MGSolver}.

\paragraph{Boundary conditions}
The boundary $\Gamma_{\delta x}$ of the discretised domain is formed by the outer faces of \emph{near-boundary} cells.
There, the following BCs can be used for the electric potential:
\begin{itemize}
   \item
         Neumann BCs $\restr{\frac{\partial \Phi}{\partial \vec{n}} }{\Gamma_{\delta x}} = g_n$ define the
         gradient---or electric field---at the boundary in normal direction.
         Since this is the flux across a cell surface, these BCs are naturally treated in finite volume schemes.
   \item
         Dirichlet BCs $\restr{\Phi}{\Gamma_{\delta x}} = g_d$ impose an electric potential at the boundary.
         Since unknowns are defined at cell centers, boundary values are extrapolated to \emph{ghost cells},
         \ie{}, cells outside the physical domain.
   \item
         Periodic BCs cyclically extend the domain.
         They are realised algorithmically by accessing unknowns in cells at opposite sides of the domain.
\end{itemize}

Neumann and Dirichlet BCs for the cell-centered discretisation are incorporated in the stencil and the RHS at near-boundary cells.
This is demonstrated for one dimension and spacing $\delta x=1$.
Consider an arbitrary 3-point stencil $\begin{bmatrix} \alpha & ~\beta & ~\gamma \end{bmatrix}$ applied at a cell $i$:
$\alpha \, \Phi_{i-1} + \beta \, \Phi_{i} + \gamma \, \Phi_{i+1} = f_i.$

For Dirichlet BCs, the boundary value $\restr{\Phi_i}{i=\frac{1}{2}} = g_d$ at cell $i=1$ is linearly extrapolated to the ghost cell, yielding $\Phi_{0} = 2 g_d - \Phi_1$.
Inserting this into the above discretisation, one obtains
the stencil $\left[ {0} \; \left( \beta - \alpha \right) \; \gamma \right]$ and the RHS $f_1  - 2 \alpha g_d$.

For Neumann BCs, the value $\restr{\frac{\partial \Phi_i}{\partial \vec{n}} }{i=\frac{1}{2}} = g_n$
is approximated by central differences at $i=\frac{1}{2}$
as $\Phi_0 - \Phi_1 = g_n$.
Substituting this in the finite volume scheme, one obtains the stencil
$\left[ {0} \; \left( \beta + \alpha \right) \; \gamma \right]$ and the RHS $f_1  - \alpha g_n$.

This treatment is preferred over the use of \emph{ghost values}
because it eliminates additional degrees of freedom.
Moreover, ghost values in the BC treatment of iterative solvers depend on values of a previous iteration,
whereas our method implicitly uses the new values for the BCs.

\paragraph{Coulomb force}
The electrostatic force acting on a particle is computed from the portions acting on each cell $b$ of a rigid body
\begin{equation}
   \vec{F}_{e} = - \delta x^3 \sum\limits_{b} \gradOp \Phi(\vec{x}_b) \; \rho(\vec{x}_b).
   \label{Eq:ElectrostatForce}
\end{equation}
The gradient of the electric potential is computed by means of finite differences that provide $\mathcal{O}({\delta x}^2)$ accuracy.
Where possible, an isotropy-preserving D3Q19 stencil is used (\cf{} \cite{2013:Ramadugu:LatticeDiffOp})
instead of a D3Q7 stencil.
With the LBM D3Q19 stencil, the gradient can be computed using $w_q$-weighted differences of neighbouring values in 18 directions $\vec{e}_q$ as
\begin{equation}
   \gradOp \Phi(\vec{x}_b) \approx \frac{1}{w_1} \sum\limits_{q=2}^{19} w_q \, \Phi(\vec{x}_b + \vec{e}_q ) \cdot \frac{\vec{e}_q }{\delta x^2}
\end{equation}

\paragraph{Subsampling}
When setting the RHS or computing the electrostatic force on the particles, the charge density of a particle at a cell is required.
The na\"{\i}ve approach is to divide the particle charge by the particle volume and to assign the resulting value to each cell whose center is overlapped by the particle.
However, due to the volume mapping, fluctuations occur and the actual charge of a particle may not be accurately represented.
This leads to errors in the computations of electric potential and electrostatic force.
Using a correction factor to adapt the charge density to the staircase-approximated volume
is inefficient in parallel computations, as it requires additional communication.
As a local alternative, we introduce a subsampling technique that computes the volume overlap ratio of the particle for each cell.
It equidistantly subdivides a cell and counts the overlapped subvolumes.
Usually used subsampling factors and their effects are provided in~\Sect{SubSec:ElPotForceValid}.

%% file: mg_solver.tex
\section{Cell-Centered Multigrid Method \label{sec:MGSolver}}
Multigrid methods are highly efficient iterative algorithms for solving large linear systems of equations.
They are based on a hierarchy of grids with different resolution levels.
On each level, high-frequency components of the error w.r.t. the resolution are efficiently eliminated by a smoothing method.
The smoothed error on a coarser grid can then be used to correct the solution on a finer grid.
Between the grid levels, information is transferred by means of restriction (fine to coarse) and prolongation
(vice versa).
Coarsening is applied recursively, with a decreasing number of unknowns on coarse grids, which significantly reduces the computational effort.
For the considered class of problems, only a fixed number of MG iterations is required to achieve a given accuracy.
This leads to a time-to-solution linear in the number of unknowns.
For a general review of MG methods we refer to~\cite{Brandt77,Trottenberg:2001:MGR}.

The geometric MG solver deployed in this paper has been developed for cell-centered discretisations
because this permits a mesh setup conforming to the LBM.
Consequently, we can seamlessly use the data structures and parallel communication routines of \Walberla{} (see \Sect{subsec:WaLBerla}).
Additionally, cell-centered discretisations lead to simpler coarsening schemes.
They allow coarsening as long as the total number of cells per dimension on a level is even 
and the number of cells at the coarsest level for each block is a multiple of two.
Node-based discretisations would require an odd number of cells and the introduction of additional nodes per block for coarsening.
A review of cell-centered MG can be found in~\cite{Mohr:2004:CCMR}.
The coarsening scheme is chosen such that
complex boundary conditions can be handled and
a seven-point stencil results on all grid levels for the Poisson problem.
The resulting solver is extensible for discontinuous dielectricity coefficients.
This can be accomplished by Galerkin
coarsening~\cite{Alcouffe:1980},
with transfer operators based on averaging restriction and a nearest neighbour prolongation.

Since we violate the condition $m_p + m_r > M$ for the polynomial orders of
prolongation $m_p$ and restriction $m_r$, where $M$ is the order of the differential operator,
we must expect that convergence rates independent of the problem size cannot be achieved~\cite{Hackbusch:1985}.
However, Braess~\cite{Braess:1995:TowardsAMG} describes that the method can be efficient for
the Poisson equation if an additional factor of about
$2$ is applied to {\em magnify} the coarse grid correction, \ie{},~we use an over-relaxation in the correction step.
A more detailed analysis is given in~\cite{Bramble:1996}.
We use this variant of cell-centered MG because it preserves the seven-point stencil on the coarse grids for Galerkin coarsening.
This keeps the method simple and avoids parallel communication overheads.
In the current implementation, V-cycles are applied.
V(1,1)-cycles do not converge reliably in our test setup,
but a higher number of pre- and post-smoothing steps leads to fast convergence.
In~\Sect{Sec:PerfMeas} V(3,3)-cycles are used, and more details will be presented.
The parallel cell-centered MG method is described in more detail in~\cite{bartuschat:2012:parallel}.

%% file: walberla_bc.tex
\section{Algorithm for Charged Particles in Fluids}
\label{Sec:OverAlg}
In this section, we present an overview of the coupled multiphysics simulation 
algorithm\footnotemark[3]\footnotetext[3]{An earlier version of this method has been presented in~\cite{bartuschat:2012:parallel}}.
The coupled physical effects are illustrated in~\Fig{fig:CP_Multiphysics_Interactions}.
\Alg{alg:CP_mine} computes the motion of the charged particles in the fluid,
including their interaction with other particles and charged planes.
Different example setups are described in~\Sect{sec:ModelProblem}.

\begin{algorithm}[h!t]
\caption{Charged Particles Algorithm.}
   \ForEach{time step}
   {
      {\color{blue} // solve Poisson problem with particle charges:}\\
         set RHS of Poisson's equation \\
      \While{residual $\geq$ tol}
      {
         perform MG V-cycle to solve the Poisson equation
      }
      {\color{blue} // solve lattice Boltzmann equation considering\\ // particle velocities:}\\
      \Begin {
         perform stream step \\
         compute macroscopic variables \\
         perform collide step \\
      }
      {\color{blue} // couple potential solver and LBM to \pe{}:}\\
      \Begin {
         apply hydrodynamic force to particles \\
         perform lubrication correction \\
         apply electrostatic force to particles\\
         \pe{} moves particles depending on forces\\
      }
   }
\label{alg:CP_mine}
\end{algorithm}

In each time step, the Poisson problem for the electrostatic potential is solved.
For this, the RHS is set depending on the particle charge densities and adapted to the BCs as described in \Sect{SubSec:ElPot}.
Then, the multigrid cycles are performed until the termination criterion is met, \ie{}, the residual $L_2$ norm is sufficiently small. 
This includes MPI communication of unknowns to \emph{ghost layers} of neighbouring processes for each level.

The LBM is performed as a fused stream-collide step (see \Sect{SubSec:TRT}).
The PDFs are communicated via MPI, and BCs including moving objects are treated.
Then the stream step and collide step are executed as described in \Sect{SubSec:TRT}.
Here the macroscopic velocity and density are required to compute the equilibrium distribution.

The hydrodynamic forces on the particles
and electrostatic forces are computed as described in \Sect{SubSec:FPI} and \Sect{SubSec:ElPot},
respectively.
For particles in close proximity, the {\em lubrication correction}
of \Sect{SubSec:LubrTheory} is applied.
All external forces 
are summed for all particles, so the time step for integrating the particle trajectories can be 
performed by invoking the \pe{} rigid body dynamics engine.
The \pe{} employs the parallel fast frictional dynamics algorithm~\cite{iglberger:2010:Pe} that
computes the new particle positions by translational and rotational motion and also resolves
rigid body collisions, taking into account friction.

\section{Implementation in WaLBerla for Multiphysics Simulations}
\Walberla{} (\emph{w}idely \emph{a}pplicable \emph{L}attice-\emph{B}oltzmann from \emph{Erla}ngen) is
a software framework for massively parallel fluid simulations that implements the methods and algorithms described above.
Due to its modular design, it can be flexibly extended~\cite{Feichtinger2011105},
and is particularly suitable for coupled multiphysics simulations.
The coupling strategy is based on accessing mutually dependent data structures
(\eg{} setting RHS of \Eqn{Eq:Poisson} dependent on particles),
and on a common BC handling concept (see~\Sect{SubSec:BCHandl}).

In this section, we present
an overview of the software concepts underlying \Walberla{}
and extend them to further support the development of multiphysics scenarios.
To this end, a novel approach to handle boundary conditions and a new
solver module are integrated into \Walberla{}.

\subsection{WaLBerla Concepts\label{subsec:WaLBerla}}
The software structure of \Walberla{} comprises a \emph{core} for sequence control and so-called \emph{applications}
that assemble functionality (\ie{} callable objects) from existing \emph{modules}.
The core initialises data structures,
performs the time stepping,
and finalises the simulation.
The central construct for time-dependent simulations is the \emph{timeloop}
that iterates through the time steps and executes the callable objects 
as they are specified by the application.

The \Walberla{} {\em functionality management}~\cite{Feichtinger2011105}
is based on dynamic application switches
that select functionality
for a given hardware at runtime.
The most fundamental callable objects are called \emph{kernels}.
For different architectures, specific optimised kernels are provided.

\Walberla{} provides a communication module using MPI for distributed memory parallelisation.
Communication is specialised for simulations on uniform grids
and supports various communication patterns.
Moveover, \Walberla{} conquers complexity by partitioning: it decomposes the domain spatially into \emph{blocks} and
functionality into \emph{sweeps}. For more details, see~\cite{bartuschat:2012:parallel}.
\paragraph{Blocks}
We use uniform rectangular grids
for discretisation with cubic cells.
Equally sized blocks of cells
are distributed to the
compute nodes.
For parallelisation and the BC handling,
a layer of ghost cells is introduced for each block.
Block data include also logistic information, such as \eg{} the location of the block within the domain or the MPI rank of the block.
The blocks allow for a {\em heterogeneous} parallel execution, 
\ie{}, the code runs on clusters that consist of different types of nodes, by invoking
architecture-specific kernels for each node.

\paragraph{Sweeps}
The time stepping within a simulation or the repeated grid traversal
of an iterative solver are organised via sweeps.
Sweeps can be performed concurrently in parallel between different blocks.
They can be concatenated and are then executed in the specified order.
Nested sweeps support iterative solvers (see \Sect{SubSec:LSEsolverModule}).
In a MG setting smoothing, restriction, and prolongation are typical sweeps.
Each sweep uses kernels and can have three phases:
a preparatory subsweep, the actual subsweep executing the numerical algorithm, and possibly a post-processing subsweep.
In particular, communication between neighbouring blocks via MPI or a time step of
the particle movement via the \pe{}
are implemented within the pre- or post-processing subsweeps.\\

\subsection{Handling Boundary Conditions}
\label{SubSec:BCHandl}
The handling of BCs
is motivated by the specific LBM requirements:
PDFs of ghost cells
must be  set such that the BCs will be fulfilled when streaming the PDFs 
into the domain (\cf{}~\cite{feichtinger:2012:Diss}).
Thus, no special treatment is required when the PDFs in the ghost layer are accessed
in a stream step.
This concept is extended for governing equations other than the lattice Boltzmann equation.

For handling multiple physical field data in a multiphysics scenario, \Walberla{}
must deal with the correct BCs for each governing equation.
Moreover, it must support individual boundaries for the different governing equations:
In our simulation, a particle imposes a BC for \Eqn{Eq:discrLBE_extForce}, but not for \Eqn{Eq:Poisson}.\\
Our new technique for handling BCs aims to satisfy the following criteria:
\begin{itemize}
   \item Generality -- applicability to different
   numerical methods.
   \item Modularity -- different functionality can be provided in independent modules.
   \item Flexibility -- different techniques for handling BCs can be provided.
   \item Configurability -- all BCs can be specified
   in input files.
\end{itemize}

\emph{Flags} are used to represent the state of a lattice site
and to indicate for each boundary cell which BC treatment has to be performed.
Different from statically determined BCs, this permits flexibility, \eg{}, the handling of moving boundaries.
Cells that are neighbouring to a boundary cell are indicated by a \emph{nearBC} flag and those
in the interior by a \emph{nonBC} flag.
Each type of boundary condition has an individual \emph{BC} flag.

Individual sets of \emph{nonBC}, \emph{nearBC} and \emph{BC} flags are specified for each governing equation.
Thus, the shape of the boundary and the
BCs can be specified individually for each field.
The actual boundary handling functionality is implemented in corresponding BC classes whose functions are executed when the associated \emph{BC} flag is found.

All BCs are handled such that they are fulfilled when
the corresponding cell is accessed in the subsequent sweep.
In case of periodic BCs, MPI processes are configured in a periodic arrangement in the \Walberla{} setup phase.
Near boundary values are then copied to the periodic neighbour's ghost layer by communication functions.
All other BCs can be handled by either direct or direction-dependent BC treatment:
Direct BC treatment directly sets the BC value at a boundary cell (as for Dirichlet BCs in node-based discretisations).
Direction dependent BC treatment sets a BC value at a boundary cell depending on the value at a neighbouring cell,
as for LBM no-slip BCs (\Eqn{Eq:BBBC}), or the BCs in cell-centered discretisations from~\Sect{SubSec:ElPot} whose handling is described in \Sect{SubSec:LSEsolverModule}.

\subsection{Linear Systems Solver Module}
\label{SubSec:LSEsolverModule}
The large sparse linear systems of equations that arise in~\Sect{SubSec:ElPot}
from the discretisation of the potential equation
can be solved in \Walberla{} by means of the \emph{lse\_solver} module that
has been designed as an efficient and robust black-box solver
on block-structured grids.
For high performance,
all operations are implemented as \emph{compact stencil} operations.
They are seamlessly integrated into \Walberla{} to avoid the overhead of using an external library.

The application
sets up the system matrix, right-hand side,
and BC handling.
When the solver sweep is added to the timeloop,
the iteration is executed.
An iterative solver requires a nested sweep that is executed until a specific convergence criterion is satisfied.

For the implicit boundary treatment and MG with Galerkin coarsening, spatially varying stencils are stored for each unknown.
In order to
reduce memory transfer, \emph{quasi-constant} stencils were introduced:
Constant stencils are used for equations with constant coefficients on the finest level.
Only at \emph{nearBC} cells, the stencils may vary for each cell and
need to be loaded individually.

\paragraph{BC Handling for Scalar Potentials}
The solver module
uses and initiates the implicit BC handling from~\Sect{SubSec:ElPot}.
The stencils are
constructed in the initialisation phase to incorporate the BCs.
For fixed geometries, when the
boundaries do not move in time, this has to be performed only once.
At the beginning of the solver sweep, the RHS is adapted to the BCs, independent of the actually used solver.
For the MG, this is only necessary on the finest grid.

At a given cell,
the stencil and RHS adaption can be applied independently for each direction,
allowing different kinds of BCs in different directions, \eg{}, at a corner.
Here we use a direction-dependent BC treatment:
the stencil entry in direction of the boundary is moved to the central entry,
and the BC value in that direction is brought to the RHS, both with a factor depending on the BC type.

\paragraph{Multigrid solver}
When the multigrid solver is applied, it automatically allocates all coarse grid data structures
and generates the coarse grid system by Galerkin coarsening (see \Sect{sec:MGSolver}).
Pre- and post-smoothing is performed with a Red-Black Gauss-Seidel algorithm that enables straightforward parallelisation by rearranging
the unknowns in a checker-board manner to decouple data dependencies.
The problem on the coarsest grid is currently solved by a conjugate gradient (CG) solver.
CG methods are more efficient than the Red-Black Gauss-Seidel, but still have non-optimal complexity.
Nevertheless, the CG is an adequate choice since the problem size on the coarsest grid remains moderate even for high levels of parallelism.

\subsection{Lubrication Correction Implementation}
Based on the description in \Sect{SubSec:LubrTheory}, a sweep has been implemented to perform the lubrication correction.
It can be added to any moving obstacle algorithm in \Walberla{} after the hydrodynamic forces have been computed.
The lubrication correction term~\Eqn{Eq:NormalLubrCorr} is added to the hydrodynamic force acting on each particle:
The algorithm iterates over all non-fixed particles residing on the current process.
In order to obtain the correction term depending on neighbouring particles or walls,
an inner loop iterates over all other objects on the current process to detect pairs.
For positive gap sizes that are smaller than the threshold $h_c$, 
the correction term is computed, incorporating the limiters described in~\Sect{SubSec:LubrTheory}.

This algorithm, as implemented, has a complexity of $\mathcal O(N^2)$ for $N$ particles. 
However, since $N$ is usually many orders of magnitude smaller than the number of fluid cells, this has negligible impact on performance. 

%% file: validation.tex
\section{Physical Validation \label{sec:Validation}}
This section presents quantitative validation results for the different components of the overall multiphysics algorithm.
To validate the fluid-structure interaction at low Reynolds numbers,
we first validate the drag force $\vec{F}_d$ acting on fixed spheres arranged in a periodic array.
We investigate the influence of volume mapping errors for different sphere radii.
The corresponding small fluctuations can also be seen in the validation of lubrication forces.
Finally, the electric potential simulation for charged particles is validated,
followed by a validation of the resulting electrostatic forces acting on the particles.

\subsection{Drag Force}
\label{SubSec:DragForceValid}
We validate the drag force $\vec{F}_d$ on spheres in an infinitely large domain.
This approach prevents wall effects that arise for alternative drag force validations with a sphere in a channel of finite width
with free-slip BCs, compared to the analytical solution in an infinitely large domain.

For equal-sized, fixed spheres arranged in a regular periodic array in Stokes flow,
an analytical solution for the dimensionless drag $K$ is derived in~\cite{sangani1982slow}.
This extends earlier results in Hasimoto~\cite{hasimoto1959periodic} for the drag on spheres in incompressible Stokes flow
for higher solid volume fractions $\phi_s$ and different particle arrangements.

$K = \frac{F_d}{6 \pi \rho_f \nu \bar{u} R}$ is the drag force on a sphere in a periodic array
immersed in an incompressible viscous fluid with average flow velocity $\bar{u}$,
normalized w.r.t the drag force on an isolated sphere.
In~\cite{sangani1982slow}, $K$ is computed for different values of $\chi = \sqrt[3]{ \phi_s/\phi_{s,max} }$,
where $\phi_{s,\max}$ is the maximal solid volume fraction for a given packing.
These results are shown to agree with values from literature within 0.5\%.
Moreover, a representation of the dimensionless drag is given as a power series in $\chi$
that allows to compute $K$ for a wider range of $\phi_s$ than given in the paper.
The relative error of this formulation w.r.t. the computed values are found to be below 1\% for $\chi < 0.95$.

For validating our algorithm, we consider the simple cubic packing, where a fixed sphere is located at the center of a cubic cell with edge length $L$.
In that case $\phi_{s,\max} = \frac{\pi}{6}$ and $\phi_s = \frac{4 \pi R^3}{3 L^3}$ for spheres of radius $R$.\\

The fluid around the sphere is simulated with the incompressible TRT model described in~\Sect{SubSec:TRT}.
In all directions periodic BCs are applied.
The fluid is driven by a uniform acceleration $g_z$ in $z$-direction by means of a body-force term for~\Eqn{Eq:discrLBE_extForce}
that is added after the stream-collide step.
We use the forcing term $F_q$ according to Luo~\cite{1998LuoForceTerm} as 
\begin{equation}
  F_{q} = w_{q} \left[\frac{(\vec{c}_{q}-\vec{u})}{c_{s}^2} + \frac{(\vec{c}_{q}\cdot\vec{u})}{c_{s}^4}\vec{c}_{q}\right] \cdot \vec{f}_\text{ext},
  \label{Eq:LBE_forcetermGuo}
\end{equation}
where $\vec{f}_\text{ext} = \rho_f g_z$ is the external force-density.
Since $F_q$ affects the momentum density, the resulting macroscopic fluid velocity is given by~\cite{ginzburg2008two}
\begin{equation}
   \vec{u} = \frac{1}{\rho_0} \left( \sum_q f_q \vec{c}_{q} + \frac{\delta t}{2} \vec{f}_\text{ext} \right).
   \label{Eq:modified_velocity}
\end{equation}

For the drag force validation, we compare the drag $K^*$ from the simulation to the reference values of $K$.
For that purpose, the hydrodynamic force $\vec{F}^*$ on the sphere is computed in the simulation according to~\Eqn{Eq:LBMhydrodynForce}.
$K$ takes into account the mean pressure-gradient~\cite{hasimoto1959periodic}.
Our simulations, however, do not include the corresponding pressure force $\vec{F}_p$ from the fluid.
Thus, the hydrodynamic force must be corrected by adding
${\vec{F}_p = \rho_f g_z V_\text{sph},}$ 
the force caused by a pressure gradient equivalent to $\rho_f g_z$. 

To compute ${K^* = \frac{F^*_L + F_{p,L}}{6 \pi \rho_{f,L} \nu_L \bar{u}^*_L R_L}}$ from the simulation results, the average fluid velocity $\bar{u}^*$
in direction of the external forcing is computed over all fluid cells~$\vec{x}_f$ as\\
${\bar{u}^* = \frac{1}{L^3} \sum_{\vec{x}_f} u_z(\vec{x})}$.
Each simulation is run until a steady state is reached, \ie{}, the relative change of $\bar{u}^*$ between two time steps is close to machine accuracy.
From the output values $F^*_L$ and $\bar{u}^*_L$, the drag $K^*$ is computed, together with the relative error ${e_{r_K}=\frac{K^*-K}{K}}$.

Additionally, the relative volume mapping error $e_{r_V} = \frac{V_\text{sim} - V_\text{sph}}{V_\text{sph}}$
of the staircase-approximated sphere volume $V_\text{sim}$ w.r.t the theoretical volume $V_\text{sph}=\frac{4}{3} R_L^3 \pi$ is computed.
To indicate the influence of the volume mapping on the error,
the relative error $c_{r_K}$ of $K^*$ corrected by the volume mapping error is computed.
It is obtained by replacing $F^*$ with $F^*_\text{cor} = \sqrt[3]{V_\text{sph}/V_\text{sim}} \, F^*,$ 
as introduced in~\cite{Goetz:2012:Diss}, based on the linear dependence of the Stokes drag force on the sphere radius.\\

In the simulations two different values of the TRT parameter $\Lambda$ are used:
$\Lambda_\text{mid}$ (\ie{} $\lambda_o = -8  (2-\omega)/(8-\omega)$)
and $\Lambda_\text{por}$ (\ie{} $\lambda_o = - (2-\omega)$), \cf{}~\Sect{SubSec:TRT}.
The influence of $\tau$ is tested for values $1.7$ and $3$, corresponding to lattice viscosities $\nu_L=0.4$ and $\nu_L=0.83$, respectively.

Dependent on $\chi$, the sphere radii are varied.
To ensure small volume mapping errors, a cell with $L=64$ is chosen.
Consequently, the radii $R_L$ are in the range $3.2$ to $28.8$ for the values of $\chi$ shown in~\Tab{Tab:ValidPerArrSphrs}.
For simplicity, $\delta x$ and $\delta t$ are set to one.
The acceleration in $z$-direction is chosen as ${g_{z,L} = 5 \cdot 10^{-7}}$, such that simulations are performed in the Stokes regime.
The particle Reynolds numbers $\mathit{Re}_p = \frac{\bar{u} \cdot 2R}{\nu}$ have the maximal value of $0.075$ at $\chi = 0.1$.

\begin{table*}[h!bt]
\centering
\caption[]{Drag force validation for single cubic cell of size $64^3$ with $g_{L,z} = 5 \cdot 10^{-7}$ for different values of $\tau$ and $\Lambda$.
           Relative errors $e_{r_K}$ and $e_{r_K}$ marked with $^*$ are computed w.r.t. the power series values of $K$. \label{Tab:ValidPerArrSphrs}}
\resizebox{1.02\linewidth}{!}{%
\begin{tabular}{|l|l|ccccccccccc|}
\hline
 & $\chi$        & 0.1       & 0.2       & 0.3       & 0.4     & 0.5     & 0.6      & 0.7       & 0.75         & 0.8         & 0.85      & 0.9        \\
 & $R_L$         & 3.2       & 6.4       & 9.6       & 12.8    & 16      & 19.2     & 22.4      & 24           & 25.6        & 27.2      & 28.8       \\
 & $e_{r_V} [\%]$   & $-0.92$   & $-0.92$ & $-1.56$ &$-0.46$&$0.58$ &$-0.62$ & $-0.07$ & $-0.09$    & $0.06$    & $0.09$  & $-0.04$  \\
\hline
\multirow{4}{*}{\parbox{1.12cm}{$\tau=1.7$ $\Lambda_\text{mid}$}}    &
   $u^*_L$       & 4.71e-3   & 1.95e-3   & 1.06e-3   & 626e-6  & 376e-6  & 227e-6   & 128e-6    & 93.6e-6       & 66.6e-6     & 46.2e-6   & 31.2e-6    \\
 & $K^*$         & 1.154     & 1.393     & 1.704     & 2.170   & 2.885   & 3.995    & 6.061     & 7.738         & 10.19       & 13.81     & 19.32      \\
 & $e_{r_K} [\%]$     & $-0.95$ & $0.34$  & $0.23$  & $0.84$& $1.51$& $0.53$ & $0.95$  & $1.04^*$   & $1.39$    & $1.28$  & $0.85$   \\
 & $c_{r_K} [\%]$     & $-0.64$ & $0.65$  & $0.75$  & $0.99$& $1.33$& $0.72$ & $0.97$  & $1.06^*$   & $1.38$    & $1.26$  & $0.86$   \\
\hline
\multirow{4}{*}{\parbox{1.12cm}{$\tau=1.7$ $\Lambda_\text{por}$}}     &
   $u^*_L$       & 4.77e-3   & 1.96e-3   & 1.07e-3   & 628e-6  & 377e-6  & 227e-6   & 128e-6    & 93.9e-6      & 66.9e-6  & 46.4e-6   & 31.4e-6    \\
 & $K^*$         & 1.140     & 1.386     & 1.698     & 2.163   & 2.877   & 3.984    & 6.043     & 7.714        & 10.15    & 13.76     & 19.24      \\
 & $e_{r_K} [\%]$     & $-2.13$ & $-0.18$ & $-0.13$ & $0.54$& $1.23$& $0.25$ & $0.65$  & $0.72^*$   & $1.04$    & $0.90$  & $0.43$   \\
 & $c_{r_K} [\%]$     & $-1.83$ & $0.12$  & $0.38$  & $0.69$& $1.05$& $0.44$ & $0.67$  & $0.74^*$   & $1.02$    & $0.88$  & $0.44$   \\
\hline
\multirow{4}{*}{\parbox{1.12cm}{$\tau=3$ $\Lambda_\text{por}$}}       &      
   $u^*_L$       & 2.29e-3   & 941e-6    & 512e-6   & 302e-6   & 181e-6  & 109e-6   & 61.9e-6   & 45.3e-6      & 32.3e-6     & 22.5e-6   & 15.2e-6    \\
 & $K^*$         & 1.139     & 1.385     & 1.697    & 2.162    & 2.873   & 3.975    & 6.022     & 7.679        & 10.09      & 13.65      & 19.04     \\
 & $e_{r_K} [\%]$     & $-2.16$ & $-0.21$ & $-0.19$ & $0.45$& $1.10$& $0.04$ & $0.31$  & $0.27^*$   & $0.44$    & $0.10$ & $-0.63$  \\
 & $c_{r_K} [\%]$     & $-1.86$ & $0.09$  & $0.33$  & $0.60$& $0.92$& $0.23$ & $0.32$  & $0.29^*$   & $0.43$    & $0.08$  & $-0.62$   \\
\hline
\end{tabular}}
\end{table*}
All parameters and the simulation results are shown in~\Tab{Tab:ValidPerArrSphrs}.
The volume mapping error $e_{r_V}$ is low for all sphere sizes, with maximal value $-1.56\%$ for $R_L=9.6$.
Moreover, $e_{r_K}$ never exceeds $2.2\%$, which arises for $\Lambda_\text{por}$ at the smallest solid volume fraction considered.
For all other cases, $e_{r_K}$ is approximately $1\%$, with largest values for $\tau=1.7$ and $\Lambda_\text{mid}$.
Using $\Lambda_\text{por}$ there, results in a smaller error for a constant value of $\tau$.
Increasing $\tau$ to $3$ reduces the error further. The increased viscosity results in higher average velocities.
As expected, the volume mapping correction leads to relative errors $c_{r_K}$ that are always smaller than $e_{r_K}$, 
with underestimated drag 
for the smallest and in one case the largest value of $\chi$.

The results are in accordance with the drag evaluation in~\cite{bogner2014drag} for periodic regular (simple cubic) arrays of spheres,
which examines the drag for TRT with $\tau=3$ and $\Lambda_\text{mid}$.
Different from this article,~\cite{bogner2014drag} focuses on flows through arrays of randomly arranged spheres for low and moderate Reynolds numbers.

\subsection{Lubrication Correction}
We show validation results for the lubrication correction in \Walberla{} according to~\Eqn{Eq:NormalLubrCorr}.
In order to validate the sphere-sphere and sphere-wall lubrication correction, two scenarios as in~\cite{Ding:2003:ExtensionLubr} are chosen:
two equally sized spheres approaching each other with the same constant velocity~$u_\text{sph}$ and a sphere vertically approaching a wall with $u_\text{sph}$.

We compare the resulting corrected lubrication forces $F^*_{Luc}$ in normal direction obtained from LBM simulations
to the analytical solution that is presented for Stokes flow in~\cite{cox:1974:motionSuspPart}.
In case of two spheres or a sphere and a fixed wall, the force is given by
\begin{equation}
   F_n = \frac{3\pi}{2} \eta \frac{1}{h_{g} \lambda^2 } u^{rel}_{n} + \mathcal O \left(\ln \left(h_g \right) \right),
   \label{Eq:LubrForceSphrSphrWall}
\end{equation}
with local surface curvature parameter $\lambda = \frac{1}{2R_a} + \frac{1}{2R_b}$.
For comparison of the results, the forces are normalised as
\begin{equation}
  f_{Norm} = \frac{F^*_{Luc}}{4 R \eta u_\text{sph}} \approx \frac{3\pi}{4} \frac{1}{h_g \lambda},
   \label{Eq:LubrForceNormal}
\end{equation}
similar to~\cite{Ding:2003:ExtensionLubr}.
The approximation for the analytical solution is derived from~\Eqn{Eq:LubrForceSphrSphrWall} and is valid for small gaps.
\Eqn{Eq:LubrForceSphrSphrWall} holds for sphere-sphere lubrication where $u^{rel}_{n} = 2 u_\text{sph}$ and for sphere-wall lubrication where ${u^{rel}_{n} = u_\text{sph}}$. \\

All simulations are performed in the Stokes regime with fixed particle Reynolds number $\mathit{Re}_p = 0.096$.
To see the influence of $\tau$ on the results, different velocities $u_\text{sph}$ or sphere sizes are chosen to keep $\mathit{Re}_p$ constant.
TRT with only $\Lambda_\text{por}$ is used, since there is no notable difference to $\Lambda_\text{mid}$ in the results.\\
We uniformly use $h_{c} = \frac{2}{3} \delta x$ in our lubrication correction.
In~\cite{nguyen2002lubrication}, this value is found to be suitable for normal lubrication force correction using BGK with $\tau = 1$.
This value is close to $\tau = 0.93301$, where TRT and BGK yield the same results w.r.t. the wall locations~\cite{2003:Ginzburg:PhysRevE.68.066614}.
Moreover, TRT exhibits no $\tau$-dependency in this regard.
Thus, in contrast to~\cite{nguyen2002lubrication}, who use BGK, considering the effective hydrodynamic radius is not necessary.\\

\begin{figure}[htb]
  \centering
     \includegraphics[width=0.48\textwidth,bb= 14 14 740 343]{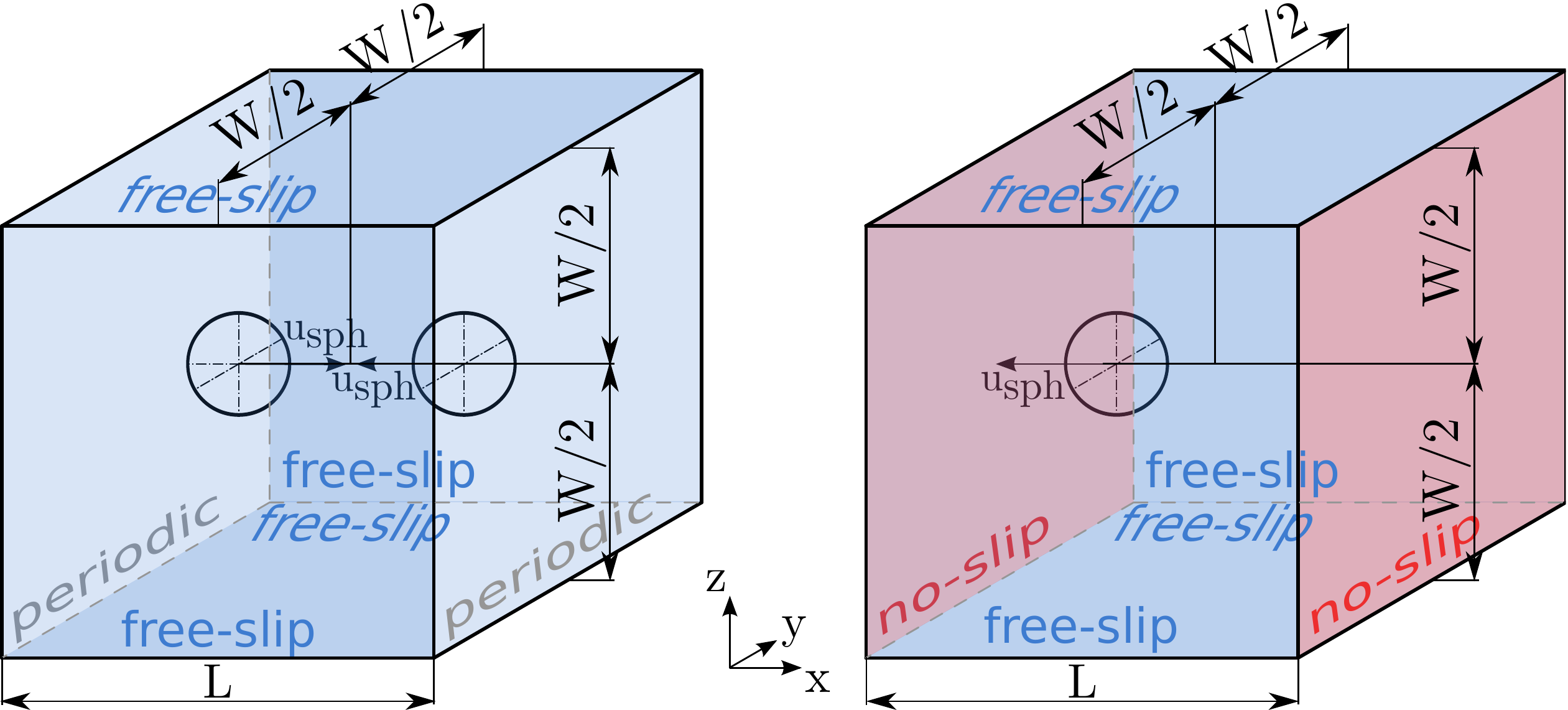}
    \caption{Setup for validation of sphere-sphere and sphere-wall lubrication correction.}
  \label{fig:LubrCorrSetup}
\end{figure}

The spheres are placed in a rectangular channel with length $L$ in $x$-direction and width $W$ in the other dimensions.
In all simulations, the spheres are centered at $W/2$ in $y$- and $z$-direction. In these directions, walls with free-slip BCs are modelled.
$W$ is chosen such that the wall effect on the force is negligible, \ie{}, about 100 times as large as the sphere diameters.
The setups are sketched in~\Fig{fig:LubrCorrSetup}.
~\\
For sphere-sphere lubrication, both spheres are initially placed with a given separation distance in $x$-direction
and are moving towards each other with a fixed constant velocity $u_\text{sph}$ each.
Here, periodic BCs are applied in $x$-direction.
($L$ is chosen such that the distance of the spheres in the periodically extended direction is at least 100 times the sphere diameter.)\\
For sphere-wall lubrication, the sphere is placed with a given distance from one of the no-slip walls in $x$-direction.
The initial distances are chosen such that the hydrodynamic force is fully developed at positions where the lubrication force is measured.
($L$ is chosen such that the distance to the wall opposite to the sphere movement is at least 100 times the sphere diameter.)\\

The sphere-sphere lubrication correction is evaluated for spheres with radius $R_L = 6$ that we frequently use in the simulations.
Since the analytical solution for the lubrication force is valid only for small gap sizes compared to the radius of curvature (see error term in~\Eqn{Eq:LubrForceSphrSphrWall}),
the numerical solution is expected to coincide for small separation distances only.
Thus, the simulations are also performed for larger spheres with radius $R_L = 48$.
The domain size for $R_L = 6$ is set to $1216 \times 1216 \times 1216 $ cells.
For ${R_L = 48}$, it is increased to $1024 \times 2048 \times 2048$ cells.
The values of $\tau$ and the corresponding lattice viscosities $\nu_L$ are shown in~\Tab{Tab:SphrSphrLubrParams}, together with the velocity of each sphere.
\begin{table}[h!b]
\centering
\caption[]{Sphere-sphere lubrication correction parameters for spheres of $R_L = 6$ and $R_L = 48$ ($\mathit{Re}_p=0.096$). \label{Tab:SphrSphrLubrParams}}
\begin{tabular}{cc|c|c}
$\tau$  & $\nu_L$ &  $u_{sph,L}$ ($R_L=6$)&  $u_{sph,L}$ ($R_L=48$)\\
\hline
1.7     & 0.4     & $1.6 \cdot 10^{-3}$   &  --                               \\
2.0     & 0.5     &  --                              &  $0.25 \cdot 10^{-3}$ \\
3.5     & 1.0     & $4.0 \cdot 10^{-3}$   &  $0.5 \cdot 10^{-3}$  \\
5.3     & 1.6     & $6.4 \cdot 10^{-3}$   &  $0.8 \cdot 10^{-3}$  \\
6.5     & 2.0     & $8.0 \cdot 10^{-3}$   &  $1.0 \cdot 10^{-3}$  \\
\end{tabular}
\end{table}
The following physical parameters are chosen:~${\rho_f=\SI{e3}{\kilo\gram\per\meter^3}}$, ${\delta x = \SI{e-3}{\meter}}$, 
and ${\delta t = \SI{0.3}{\second}}$---such that $\nu = \SI{1.3e-6}{\meter^2\per\second}$ for $\nu_L=0.4$.\\
Sphere-wall lubrication correction is validated for spheres of different size moving with constant velocity $u_{sph,L} = 10^{-3}$.
The domain size is set to $1216 \times 1216 \times 1216$ cells for $R_L=6$, $1600 \times 1792 \times 1792$ for $R_L=8$ and $R_L=9$, and $1216 \times 2432 \times 2432$ for $R_L=12$.
The sphere radii are shown in~\Tab{Tab:SphrWallLubrParams}, together with $\nu_L$ and the corresponding $\tau$.
Different from the sphere-sphere lubrication correction, we chose ${\delta t=\SI{0.125}{\second}}$---such that ${\nu=\SI{e-6}{\meter^2\per\second}}$ for $R_L=6$.
All other physical parameters are kept identical.
\begin{table}[h!]
\centering
\caption{Sphere-wall lubrication correction parameters for different sphere radii ($\mathit{Re}_p=0.096$). \label{Tab:SphrWallLubrParams}}
\begin{tabular}{l|cccc}
$R_L$       & 6       & 8        & 9         & 12     \\
\hline
$\nu_{L}$   & $1/8$   & $1/6$    & $3/16$    & $1/4$  \\
$\tau$      & $0.875$ & $1.0$    & $1.0625$  & $1.25$ \\
\end{tabular}
\end{table}

\begin{figure}[h!t]
  \centering
    \includegraphics[width=0.48\textwidth,bb=14 14 745 426]{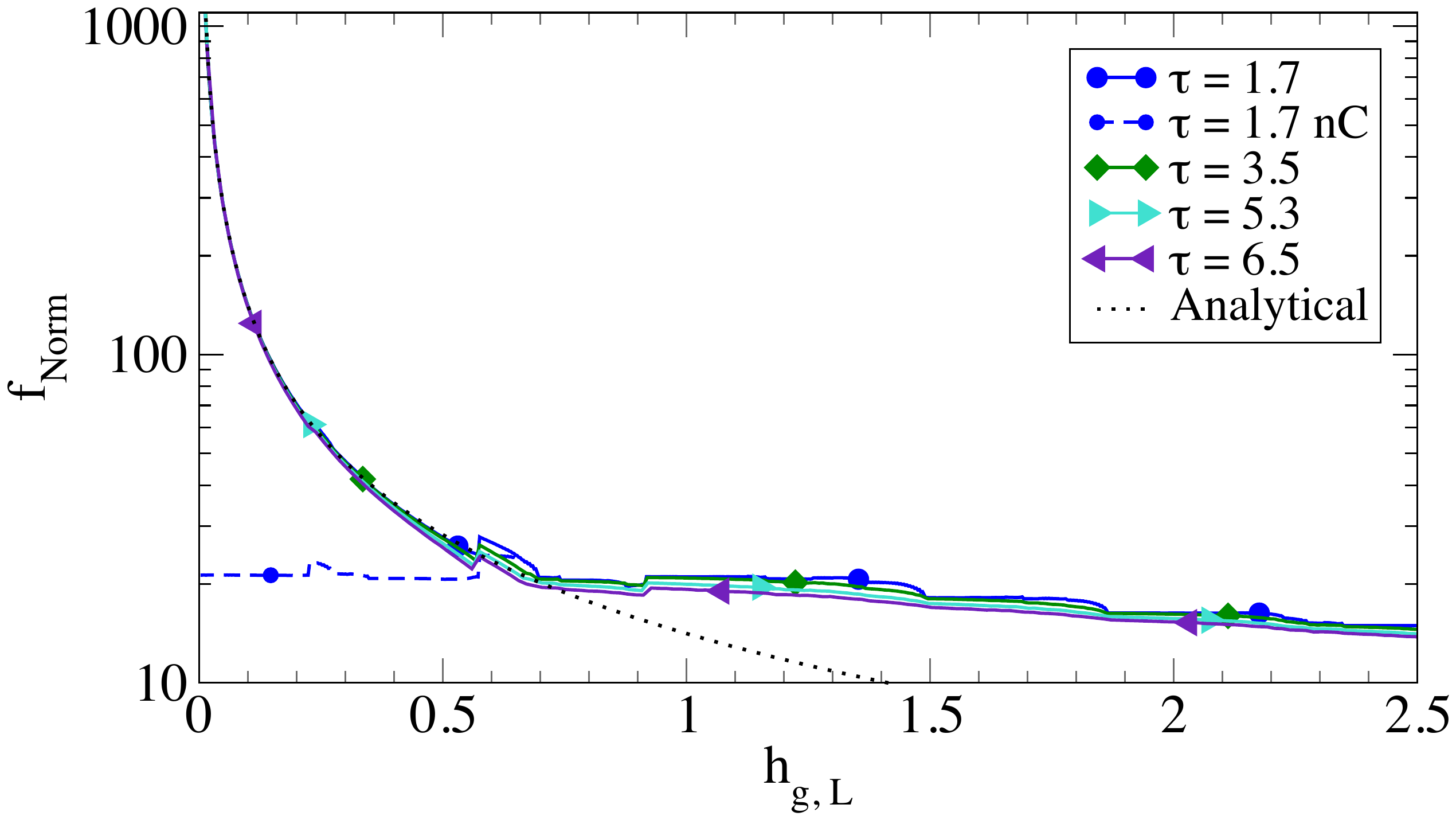}
    \caption{Validation of sphere-sphere lubrication correction for spheres of radius $R_L = 6$.}
  \label{fig:SphrSphr_6R0_LubrValidation}
\end{figure}

\begin{figure}[h!tb]
  \centering
  \includegraphics[width=0.48\textwidth,page=1]{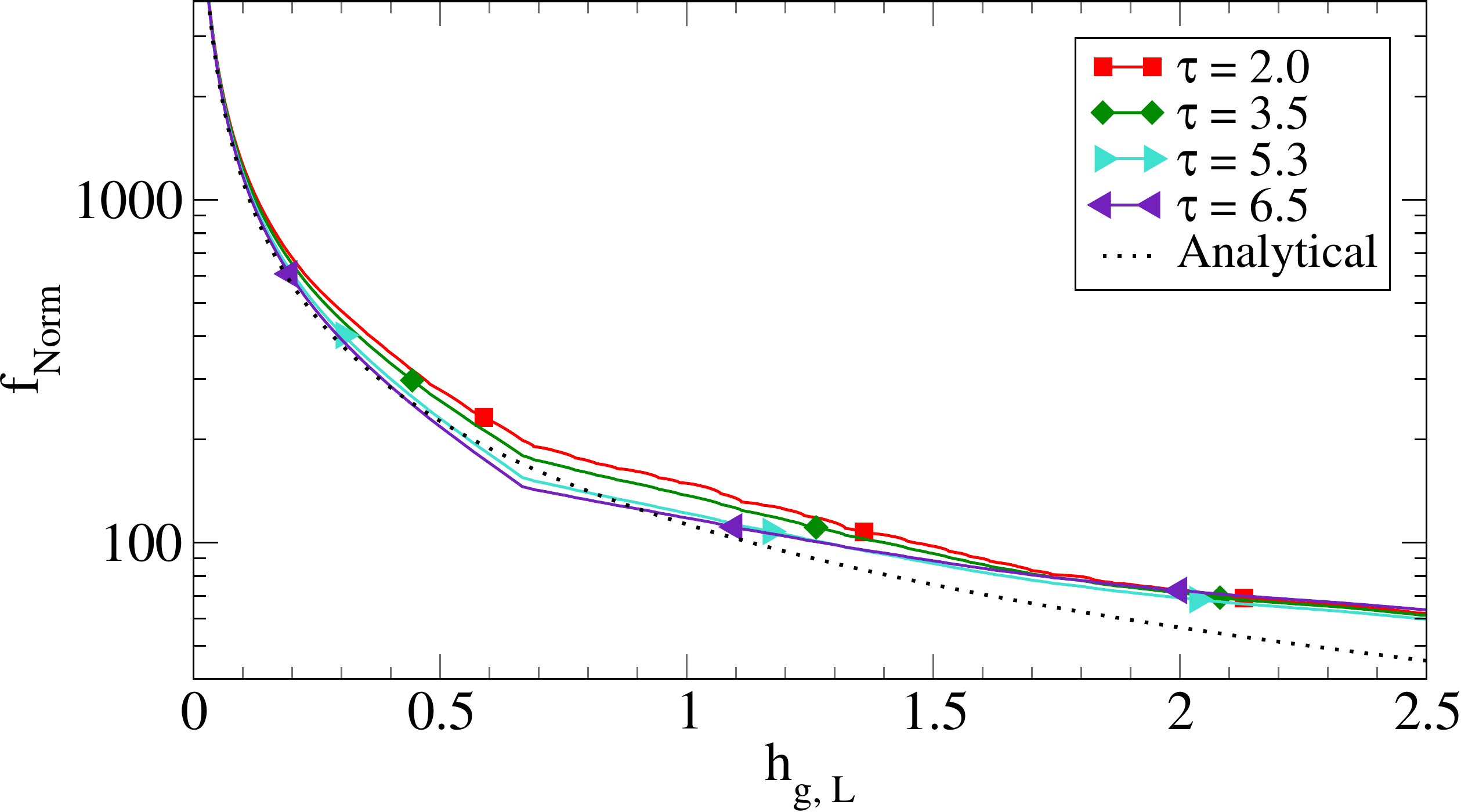}
    \caption{Validation of sphere-sphere lubrication correction for spheres of radius $R_L = 48$.}
  \label{fig:SphrSphr_48R0_LubrValidation}
\end{figure}

\begin{figure}[h!]
  \centering
  \includegraphics[width=0.48\textwidth,page=1]{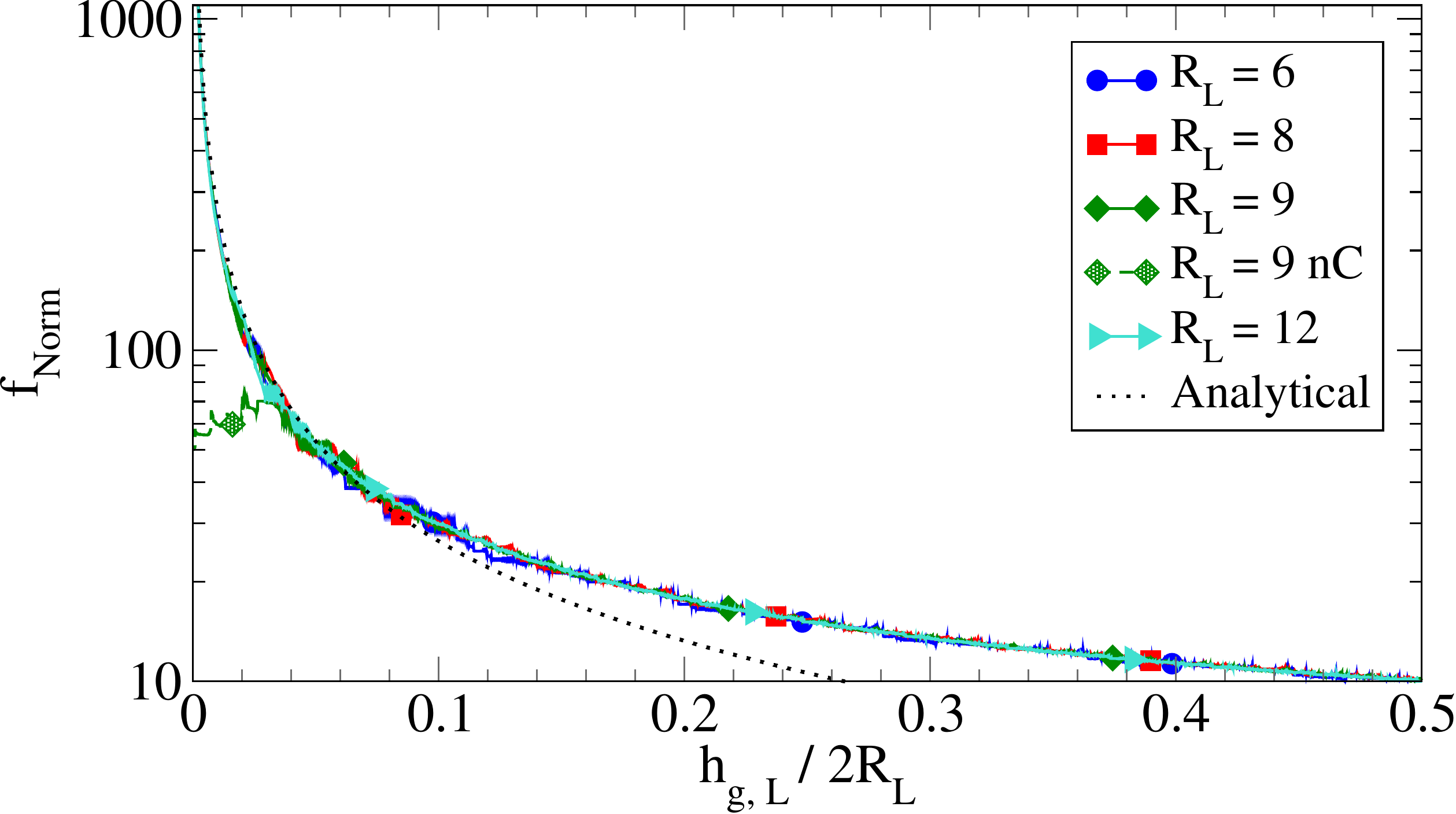}
    \caption{Validation of sphere-wall lubrication correction for different sphere sizes.}
  \label{fig:SphrWallLubrValidation}
\end{figure}

For each simulation, the computed force $F^*_{Luc}$ on a sphere including lubrication correction is normalised according to~\Eqn{Eq:LubrForceNormal}.
The results for sphere-sphere lubrication at different gap sizes $h_{g,L}$
are depicted in~\Fig{fig:SphrSphr_6R0_LubrValidation} and~\Fig{fig:SphrSphr_48R0_LubrValidation} for ${R_L = 6}$ and ${R_L = 48}$, respectively.
The results for sphere-wall lubrication are depicted in~\Fig{fig:SphrWallLubrValidation} at different normalised gap sizes $\frac{h_{g}}{2R}$.
In each figure, the normalised approximation to the analytical solution in~\Eqn{Eq:LubrForceNormal} is plotted for reference.

Additionally, the normalised hydrodynamic force
without lubrication correction is shown in~\Fig{fig:SphrSphr_6R0_LubrValidation}
and in~\Fig{fig:SphrWallLubrValidation}, indicated by `nC'. 
In~\Fig{fig:SphrSphr_6R0_LubrValidation}, it stays approximately constant below one lattice spacing,
as already reported in~\cite{Ladd_1994_PartSuspPt2}.
In~\Fig{fig:SphrWallLubrValidation}, it also stays constant for gaps smaller than one lattice spacing, but then decreases for $h_{g,L} < 0.5$. 
The reason for the decrease is that a lattice cell whose center is overlapped by the particle is no longer considered a fluid cell.
Consequently, in the cell at minimum distance between sphere and wall, the pressure is no longer represented correctly.
This leads there to an attractive force and an overall decreasing force on the sphere.
In contrast, the spheres of the sphere-sphere lubrication validation move symmetrically towards the center of the cell between the spheres,
and thus this cell center is never overlapped.

The lubrication correction method reproduces the analytical solution for decreasing gap sizes in all simulations (see \Fig{fig:SphrSphr_6R0_LubrValidation}
to \Fig{fig:SphrWallLubrValidation}).
For larger gap sizes, the forces from the simulations and the analytical solution deviate increasingly, especially for $R_L=6$ (see \Fig{fig:SphrSphr_6R0_LubrValidation}).
This difference is reduced in~\Fig{fig:SphrSphr_48R0_LubrValidation}, where the simulation results agree better with the analytical solution also for larger gap sizes.

Overall, the forces in~\Fig{fig:SphrSphr_48R0_LubrValidation} coincide very well for large and small gap sizes.
Only for $h_{g,L}$ smaller than approximately two the forces start diverging, depending on $u_\text{sph}$ and $\tau$.
Nevertheless, all simulated forces lie closely around the analytical solution in that range of $h_{g,L}$.

For the case of smaller spheres, as evaluated in~\Fig{fig:SphrSphr_6R0_LubrValidation} and~\Fig{fig:SphrWallLubrValidation},
the volume mapping error becomes visible.
Naturally, this error decreases with increasing sphere size, as indicated by the ripple in~\Fig{fig:SphrWallLubrValidation}.
The influence of $\tau$ and $u_\text{sph}$ on the force fluctuations due to volume mapping effects
for a constant sphere size can be seen in~\Fig{fig:SphrSphr_6R0_LubrValidation}, where the jumps decrease with increasing $\tau$ and $u_\text{sph}$.
In both cases, this error increases for ${h_{g,L} < 1}$, but is insignificant for small gap sizes because the lubrication correction dominates.

Overall, the forces in~\Fig{fig:SphrSphr_6R0_LubrValidation} differ only slightly.
The hydrodynamic force marginally decreases with increasing $\tau$ and $u_\text{sph}$.
Tests with higher $\mathit{Re}_p$ have shown that for a fixed $\tau$ but increasing $u_\text{sph}$,
both, the normalised hydrodynamic forces and the volume mapping errors, decrease slightly.
Thus, these effects can be mainly attributed to $u_\text{sph}$.
Also the forces in~\Fig{fig:SphrWallLubrValidation} differ only slightly.
No systematic dependence on $R_L$ and $\tau$ values could be determined.
The reason is that $u_\text{sph}$ is constant, reflecting the minor influence of $\tau$ on the force.
~\\

The lubrication force correction formula is often also applied for larger Reynolds numbers than the ones presented here.
This is justified because, due to the small gaps, one can still approximate the flow by Stokes flow~\cite{janoschek2013accurate}.

\subsection{Electric Potential and Electrostatic Force}
\label{SubSec:ElPotForceValid}
To validate the electric potential computation, we simulate a homogeneously charged sphere in a large domain.
The numerical solution is then compared to the analytical solution for the potential distribution $\Phi (\vec{r})$,
which in case of an infinitely large domain is
\begin{equation}
    \Phi (\vec{r}) = 
        \left\{
        \begin{array}{l l}
          \frac{1}{4 \pi \varepsilon} \frac{Q}{ | \vec{r} |  }                                                 & \quad \text{if } | \vec{r} | \geq R \\
          \frac{1}{4 \pi \varepsilon} \frac{Q}{2 R} \left( 3 - \left( \frac{|\vec{r}| }{ R } \right)^2 \right) & \quad \text{if }  | \vec{r} | < R,
        \end{array} 
        \right.
\end{equation}
with the distance from the sphere center $\vec{r}$, the total charge $Q$ of the sphere and its radius $R$.
Simulating a domain so large that the electric potential decays to zero at the boundary is not feasible.
Thus, we set the boundary values to the analytical solution in our validation experiments.

\begin{figure}[ht]
  \centering
  \includegraphics[width=0.48\textwidth,page=1]{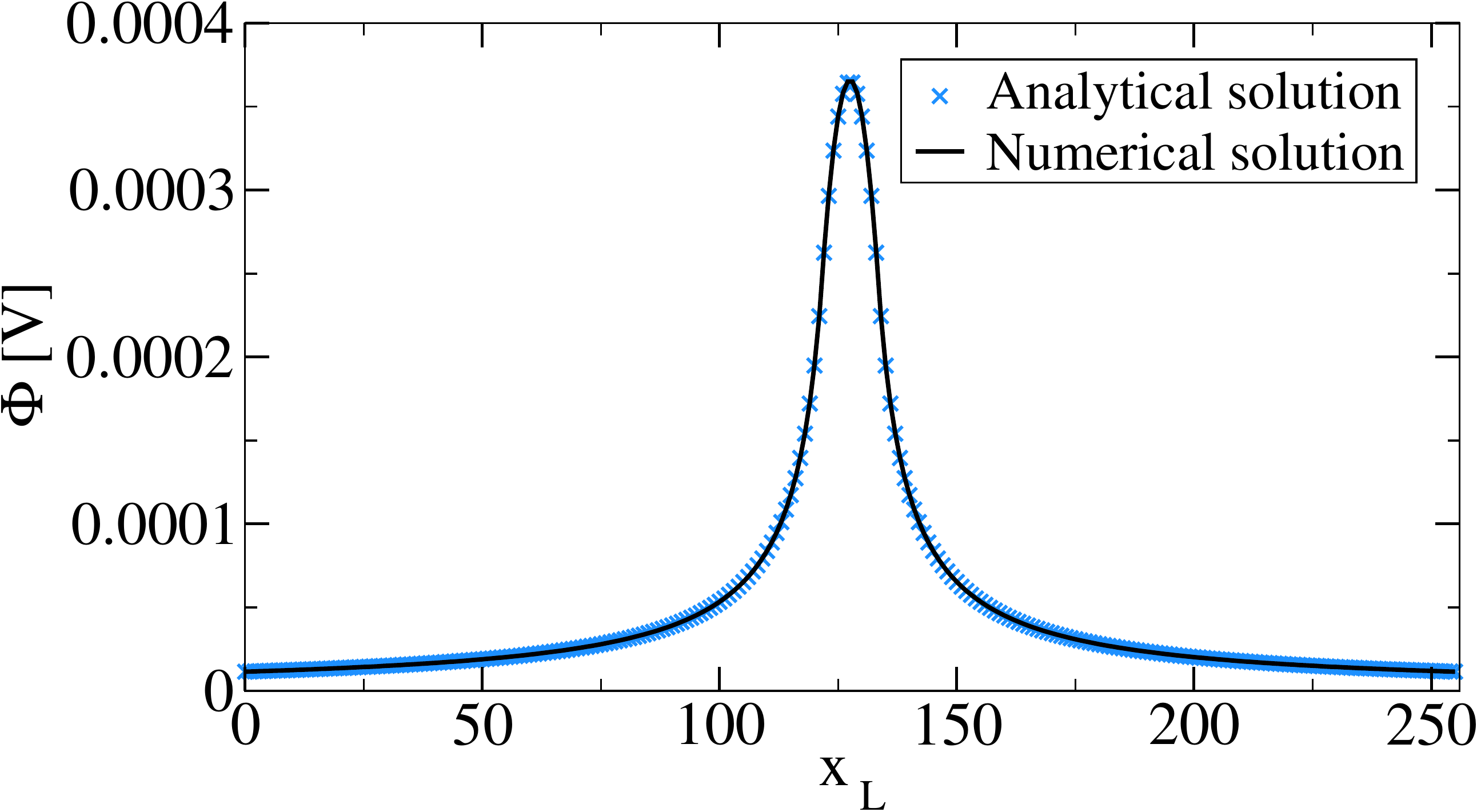}
    \caption{Analytical and numerical solution for electric potential from homogeneously charged sphere.}
  \label{Fig:PotValidation}
\end{figure}

For a sphere with radius $R_L = 6$ placed at the center of a domain of size $256^3$,
both, analytical ($\Phi$) and numerical ($\Phi^*$) solutions, are depicted in \Fig{Fig:PotValidation} along a line through the domain center.
The simulation parameters are $Q= \SI{8000}{\elementarycharge}$ and $\varepsilon=78.5\cdot \varepsilon_0$, with vacuum permittivity $\varepsilon_0$.
Additionally, a subsampling factor of three (\ie{} subdivision of each cell in three intervals for each dimension) is used. 
Clearly, both solutions are in good agreement.

A quantitative assessment of the maximal errors in the electric potential computation is performed for different particle radii and subsampling factors $s_f$.
First, the particle position with maximal volume mapping error $e_{r_V}$ (see \Sect{SubSec:DragForceValid}) is determined.
To this end, the sphere is placed at the domain center and then repeatedly shifted by a small distance in all dimensions (up to $\pm 0.51 \, \delta x / s_f$).
From the errors $e_{r_V}$, the maximal error $e_{{r_V}\text{max}}$ is obtained. Additionally, the average volume mapping error $ \overline{|e_{r_V}|}$ is computed from the absolute values of $e_{r_V}$.
For the position of $e_{{r_V}\text{max}}$, the relative error $e_{r_\Phi}=\frac{\Phi^*-\Phi}{\Phi}$ of the numerical solution $\Phi^*$ is computed in the whole domain.
In \Tab{Tab:ValidPotSubs}, the $L_2$ norm and the maximum norm of that error is shown, together with $e_{{r_V}\text{max}}$ and $\overline{|e_{r_V}|}$ for different sphere sizes and subsampling factors.\\
\begin{table}[h!]
\centering
	\caption[]{Electric potential validation for homogeneously charged sphere in domain of size $256^3$ for different radii and subsampling factors.
	Comparison of relative volume mapping errors $e_{r_V}$ and of relative electric potential errors $e_{r_\Phi}$ at the position of maximal volume mapping error.
   \label{Tab:ValidPotSubs} }
\resizebox{0.5\textwidth}{!}{
\begin{tabular}{|l|l|cccc|}
\hline
\multicolumn{2}{|c|}{\multirow{2}{*}{}}&\multicolumn{4}{c|}{pot. subsampling (${s_f}_{\Phi}$)} \\
\multicolumn{2}{|c|}{} & 1    & 2         & 3         & 4         \\
\hline
\multirow{4}{*}{\parbox{1.3cm}{$R_L=4$}}  &
   $ \overline{|e_{r_V}|} $          & 1.14\%  & 0.284\% & 0.145\% & 0.081\%        \\ 
 & $ e_{{r_V}\text{max}} $           & 7.06\%  & 2.58\%  & -1.43\% & 0.72\%         \\ 
 & $ \| e_{r_\Phi} \|_{L_2} $        & 1.87\%  & 0.622\% & 0.568\% & 0.192\%        \\ 
 & $ \| e_{r_\Phi} \|_\infty $       & 10.9\%  & 3.37\%  & -2.06\% & 1.13\%         \\ 
\hline
\multirow{4}{*}{\parbox{1.3cm}{$R_L=6$}} &
   $ \overline{|e_{r_V}|} $          & 0.638\% & 0.140\% & 0.065\% & 0.039\%       \\ 
 & $ e_{{r_V}\text{max}} $           & -2.74\% & -1.43\% & -0.33\% & -0.25\%       \\ 
 & $ \| e_{r_\Phi} \|_{L_2} $        & 0.927\% & 0.568\% & 0.290\% & 0.273\%       \\ 
 & $ \| e_{r_\Phi} \|_\infty $       & -4.48\% & -2.19\% & -0.80\% & -0.61\%       \\ 
\hline
\multirow{4}{*}{\parbox{1.3cm}{$R_L=8$}}  &
   $ \overline{|e_{r_V}|} $          & 0.285\% & 0.081\% & 0.040\% & 0.021\%       \\ 
 & $ e_{{r_V}\text{max}} $           & 2.58\%  & 0.72\%  & -0.25\% & -0.14\%       \\ 
 & $ \| e_{r_\Phi} \|_{L_2} $        & 0.622\% & 0.192\% & 0.273\% & 0.251\%       \\ 
 & $ \| e_{r_\Phi} \|_\infty $       & 3.37\%  & 0.97\%  & -0.67\% & -0.48\%       \\ 
\hline
\multirow{4}{*}{\parbox{1.3cm}{$R_L=9$}}  &
   $ \overline{|e_{r_V}|} $          & 0.258\% & 0.066\% & 0.028\% & 0.018\%       \\ 
 & $ e_{{r_V}\text{max}} $           & 1.91\%  & -0.33\% & 0.32\% & -0.11\%        \\ 
 & $ \| e_{r_\Phi} \|_{L_2} $        & 0.444\% & 0.290\% & 0.185\% & 0.244\%       \\ 
 & $ \| e_{r_\Phi} \|_\infty $       & 2.48\%  & -0.82\% & 0.44\% & -0.39\%        \\ 
\hline
\multirow{4}{*}{\parbox{1.3cm}{$R_L=12$}} &
   $ \overline{|e_{r_V}|} $          & 0.146\% & 0.040\% & 0.018\% & 0.009\%       \\ 
 & $ e_{{r_V}\text{max}} $           & -1.43\% & -0.25\% & -0.11\% & -0.10\%       \\ 
 & $ \| e_{r_\Phi} \|_{L_2} $        & 0.567\% & 0.273\% & 0.244\% & 0.243\%       \\ 
 & $ \| e_{r_\Phi} \|_\infty $       & -2.02\% & -0.58\% & -0.39\% & -0.39\%       \\ 
\hline
\end{tabular}
}
\end{table}
The average error caused by the volume mapping exceeds 1\% only for radius 4 when no subsampling is used.
For spheres with radii of at least 6, even the maximal error is below 3\% without subsampling.
Additional subsampling reduces the volume mapping further. As expected, $\overline{|e_{r_V}|}$ is approximately halved when incrementing $s_f$ by one.
The relative error of the potential over the whole domain, measured in the $L_2$ norm, is well below 2\% for all considered sphere sizes, even if no subsampling is used.
Only the maximal value of $e_{r_\Phi}$ may become higher than 1\% without subsampling.
With higher subsampling and larger spheres, all shown errors decrease until the lower bound of the discretisation error is reached.
The results also show that the errors for same $s_f R_L$ match very well.\\

The Coulomb force computation according to~\Sect{SubSec:ElPot} is validated by simulating a charged sphere in a domain with homogeneous electric field.
For different sphere sizes and subsampling factors, the sphere is placed at the position of maximal volume mapping error near the domain center.
From the simulated electrostatic force $F_C^*$ and the analytical solution $F_C$, the relative error is computed by $e_{r_{F_C}}=\frac{F_C^*-F_C}{F_C}$.
The domain size and physical properties of the spheres are the same as in the electric potential validation.
At the left and right boundary, Dirichlet conditions with different values are applied ($\Phi_W = 0$, $\Phi_E = -10V$).
In all other dimensions, homogeneous Neumann BCs are applied.
From the potential difference across the domain and the particle charge, the reference value of the Coulomb force is obtained.

The sphere is first placed at positions with maximal volume mapping error occurring for a given subsampling factor of the Coulomb force.
Then $e_{r_{F_C}}$ is equal to $e_{{r_V}\text{max}}$ for same subsampling factors of potential and electrostatic force.
For unequal subsampling factors, $e_{r_{F_C}}$ changes only slightly.
\Ie{}, the smaller volume mapping error for the electric potential computation has only minor effect on the accuracy of the force.

For the positions with $e_{{r_V}\text{max}}$ of the electric potential subsampling factor, errors are shown in~\Tab{Tab:ValidClmbForceSubs_PotMaxErr}.
\begin{table}[h!]
\centering
\caption[]{Coulomb force validation in domain of size $256^3$, relative errors $e_{r_{F_C}}$ for different radii and subsampling factors.
           Particles are located at the position of maximal volume mapping error for potential subsampling.\label{Tab:ValidClmbForceSubs_PotMaxErr} }
\begin{tabular}{|l|l|ccc|c|}
\hline
\multicolumn{2}{|c|}{} & \multicolumn{3}{c|}{force subsampling (${s_f}_{F_C}$)} &  \\
\multicolumn{2}{|c|}{} & 1          & 2         & 3         & $R_L$ \\
\cline{2-6}
\parbox{2.5mm}{\multirow{14}{*}{\parbox{1.3cm}{\rotatebox[origin=c]{90}{pot. subsampling (${s_f}_{\Phi}$)}}}} &
  1       & 7.06\%      & -0.49\%   & -0.25\% & \multirow{3}{*}{{4}} \\ 
& 2       & -5.27\%     & 2.58\%    & -0.64\% & \\ 
& 3       & -5.68\%     & 1.48\%    & -1.43\% & \\ 
\cline{2-6}
& 1       & -2.74\%     & 0.27\%    & -0.13\% & \multirow{3}{*}{{6}} \\ 
& 2       & -0.11\%     & -1.43\%   & -0.03\% & \\
& 3       & 0.18\%      & -0.41\%   & -0.33\% & \\ 
\cline{2-6}
& 1       & 2.58\%      & 0.20\%    & 0.05\%  &\multirow{3}{*}{{8}} \\ 
& 2       & -1.80\%     & 0.72\%    & -0.19\% & \\ 
& 3       & -0.44\%     & -0.04\%   & -0.25\% & \\ 
\cline{2-6}
& 1       & 1.91\%      & 0.26\%    & 0.02\%  &\multirow{3}{*}{{9}} \\ 
& 2       & 0.27\%      & -0.33\%   & 0.01\%  & \\
& 3       & 1.91\%      & 0.14\%    & 0.32\%  & \\ 
\cline{2-6}
& 1       & -1.43\%     & -0.02\%   & -0.08\% &\multirow{2}{*}{{12}} \\
& 2       & -0.08\%     & -0.25\%   & 0.02\%  & \\
& 3       & 0.09\%      & -0.01\%   & -0.11\% & \\
\hline
\end{tabular}
\end{table}
Applying the the same subsampling to potential and electrostatic force computation results in errors equal to $e_{r_V}$.
For different combinations of subsampling factors, the error decreases.
Then the position corresponds to $e_{{r_V}\text{max}}$ of the potential subsampling where $e_{r_V}$ of the force subsampling is smaller.

We conclude that subsampling is particularly important when computing the electrostatic force, as its value is directly proportional to the particle charge.
To obtain errors in the electrostatic force comparable to those of the fluid-particle interaction, a subsampling factor of two is sufficient for spheres with radius $6\delta x$ or $8\delta x$.
Spheres with smaller radius require higher subsampling and those with larger radius require no subsampling.

%% file: results.tex
\section{Towards Realistic Multiphysics Simulations\label{sec:ModelProblem}}
We proceed to present two showcases that demonstrate the proper operation of the coupled multiphysics simulation.
For the showcases, animations are available via permalinks:
\begin{enumerate}
   \item The attraction of homogeneously charged particles in fluid flow by an oppositely charged surface and their deposition on that surface\footnotemark[3]\footnotetext[3]{\url{https://www10.cs.fau.de/permalink/igahque9ie}}. \label{CP_AgglomShowcase}
   \item The separation of oppositely, homogeneously charged particles in fluid flow in a bisecting micro-channel, due to an electric field imposed by charged surfaces\footnotemark[4]\footnotetext[4]{\url{https://www10.cs.fau.de/permalink/kai4Iepha3}}. \label{CP_SeparShowcase}
\end{enumerate}

\begin{figure}[h]
  \centering
  \subfigure[Initial setup]{
    \label{fig:VidSetupInit}
  \includegraphics[width=0.22\textwidth]{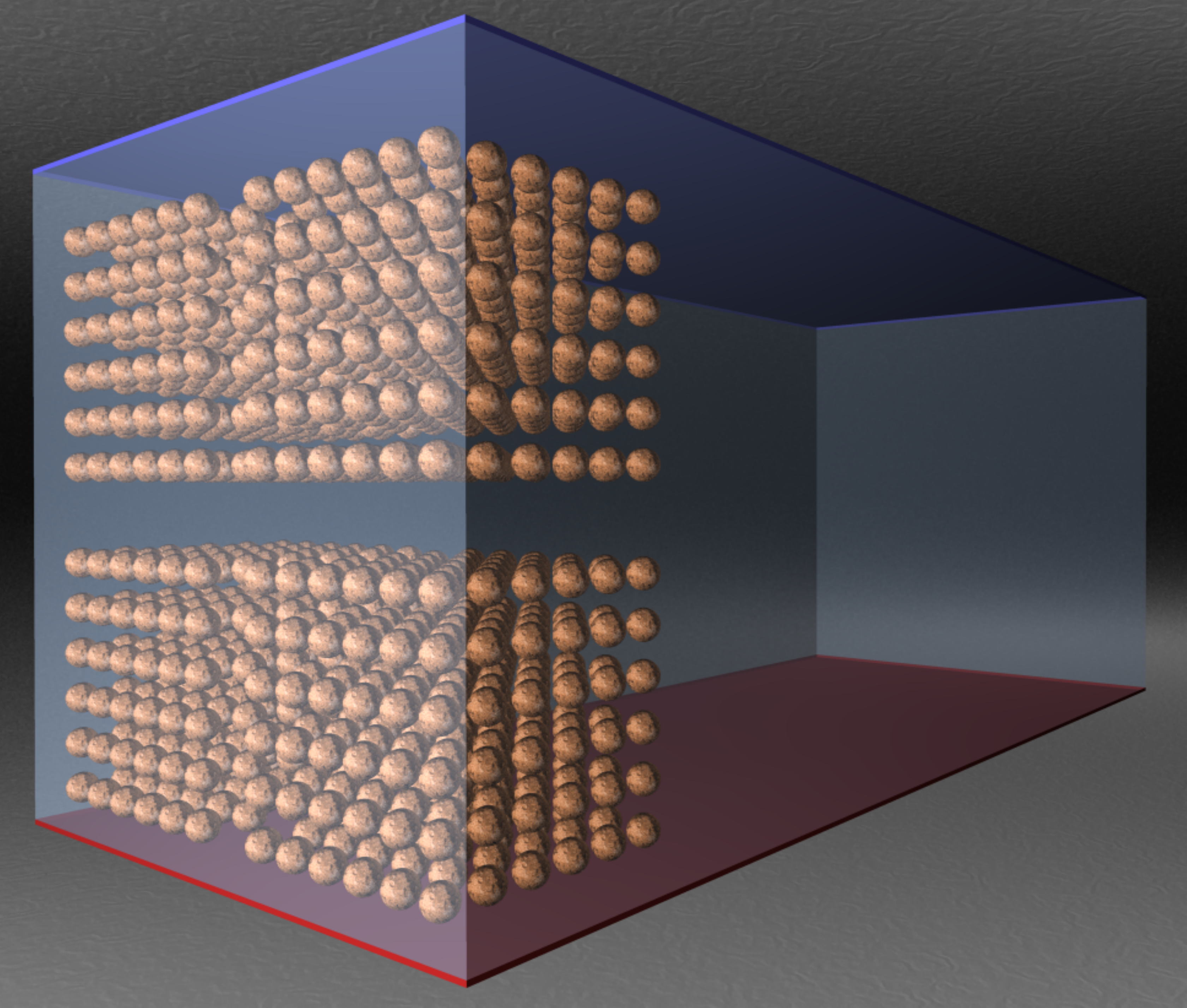}
  }
  \subfigure[Later state]{
    \label{fig:VidSetupImg167}
  \includegraphics[width=0.22\textwidth]{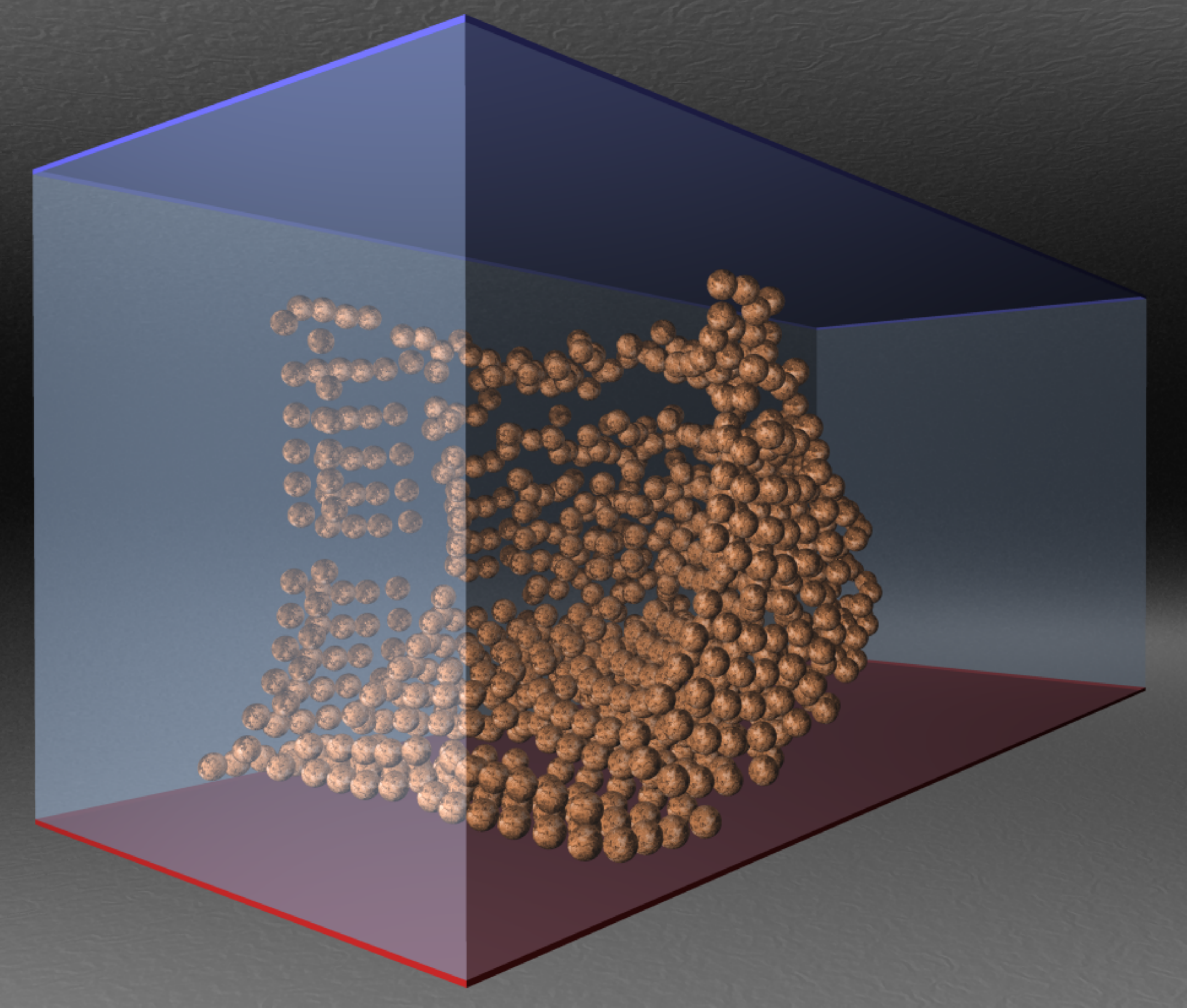}
  }
  \caption{Initial and a later state (time step \num{35200}) of the charged particle agglomeration simulation. \label{fig:VidSetup} }
\end{figure}

\begin{figure}[h!]
  \centering
  \subfigure[Particles with pathlines in time step \num{75300}]{
    \label{fig:SeparVid250PathlParticles}
  \includegraphics[width=0.36\textwidth,trim={0 0 0 6.2cm},clip]{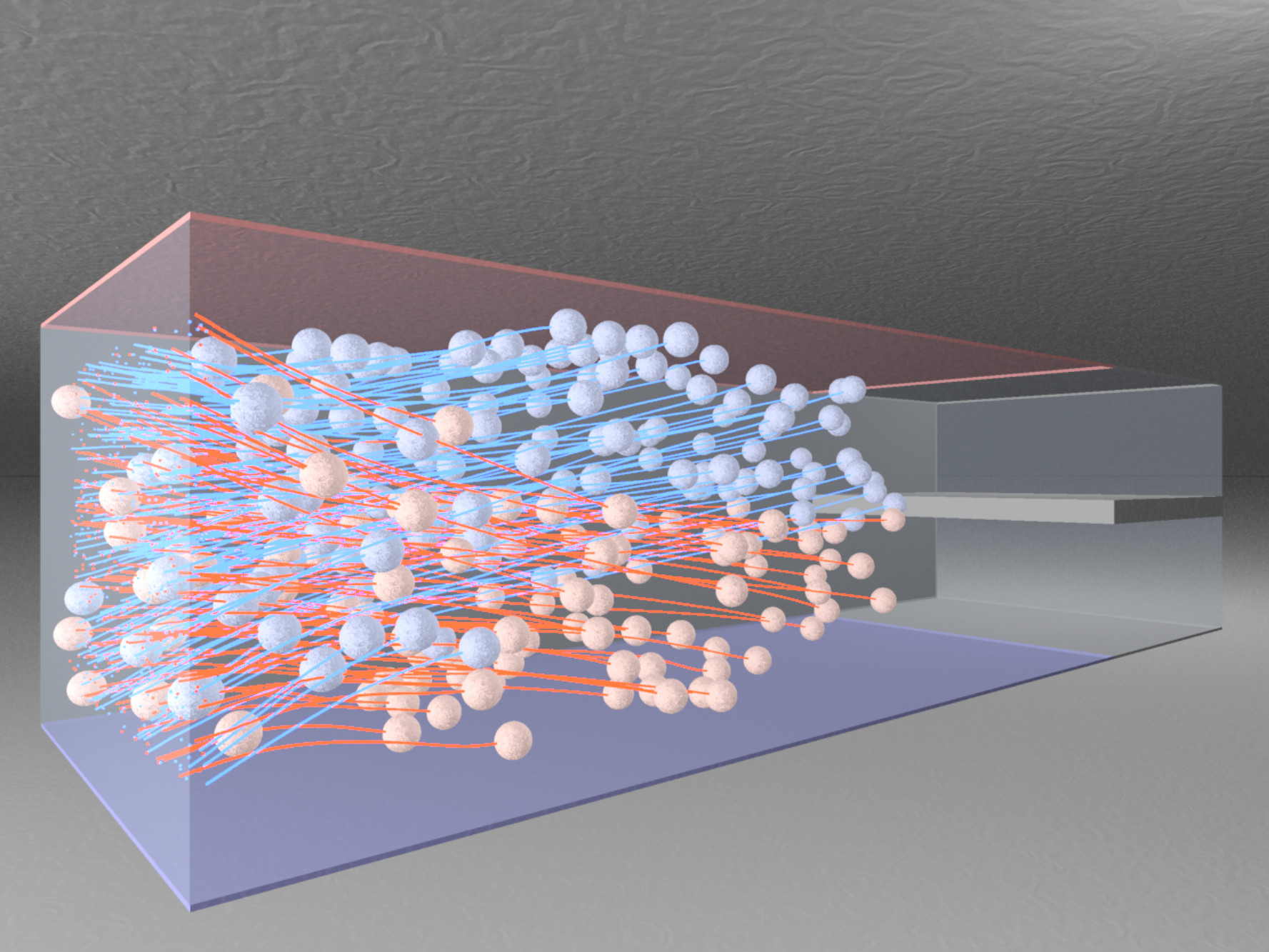}
  }\\
  \subfigure[Final pathlines after \num{210000} time steps]{
    \label{fig:SeparVidAllPathlines}
  \includegraphics[width=0.36\textwidth,trim={0 0 0 6.2cm},clip]{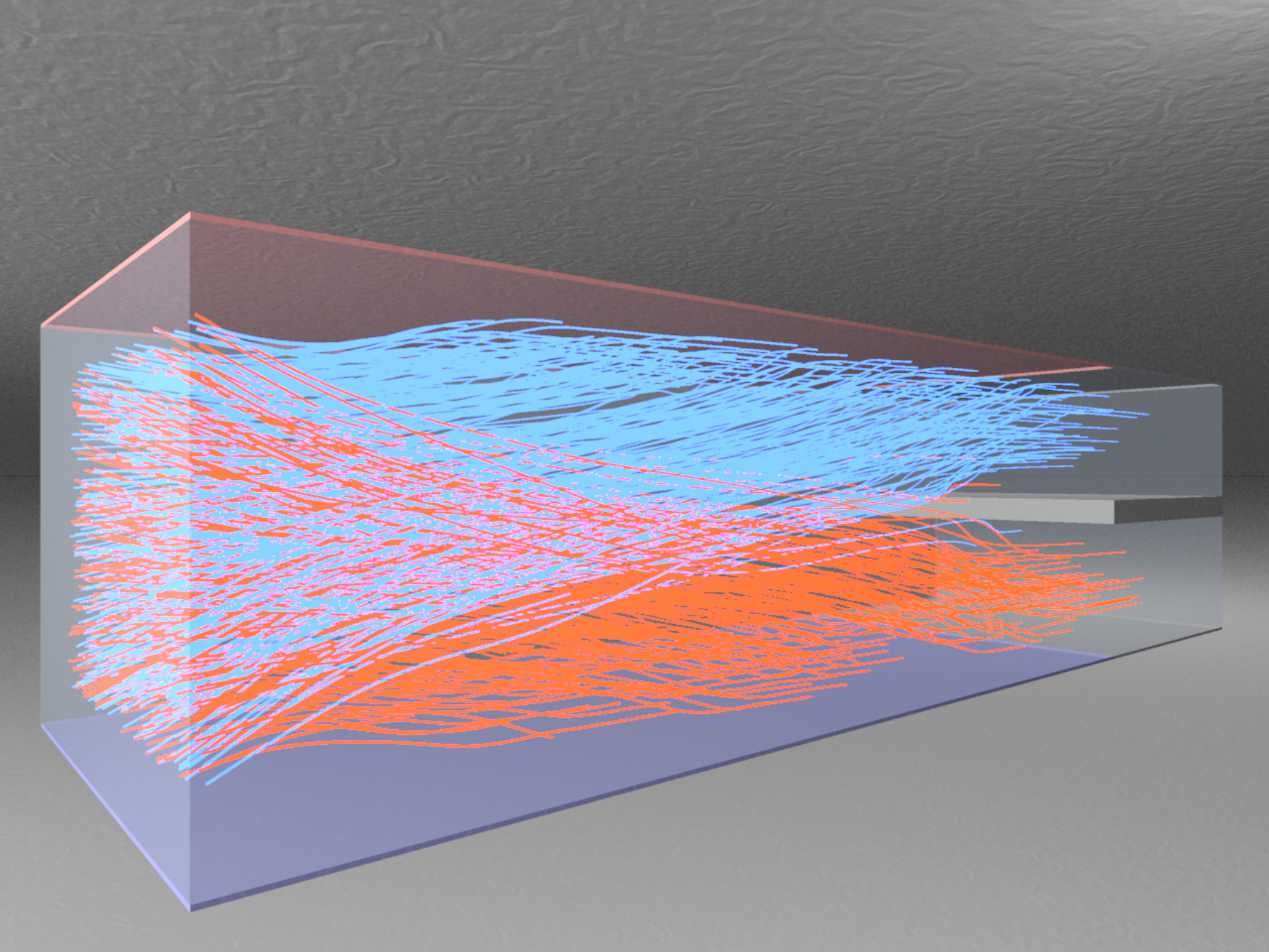}
  }
  \caption{Particle distribution and pathlines for the charged particle separation simulation. \label{fig:SeparVidSetup} }
\end{figure}

Both showcases simulate a micro-channel with fluid flow in longitudinal direction.
Showcase~\ref{CP_AgglomShowcase} with a channel size of $\SI{2.56}{\milli\meter}\times{}\SI{5.76}{\milli\meter}\times{}\SI{2.56}{\milli\meter}$ is depicted in~\Fig{fig:VidSetup}.
The flow results from an inflow BC (\Eqn{Eq:LBMMovWall}) with velocity \SI{1}{\milli\meter\per\second} at both, front and back side.
Showcase~\ref{CP_SeparShowcase} is depicted in~\Fig{fig:SeparVidSetup}. 
The channel of size $\SI{2.56}{\milli\meter}\times{}\SI{7.68}{\milli\meter}\times{}\SI{1.92}{\milli\meter}$ is split into two halves after \SI{5.76}{\milli\meter} by a fixed beam of height \SI{160}{\micro\meter} with no-slip BCs that is attached to the side walls.
The flow is imposed here by the inflow BC with velocity \SI{1}{\milli\meter\per\second} at the front side. 
At the back side zero pressure outflow is modelled with a \emph{pressure anti-bounce-back} BC from~\cite{ginzburg2008two}.
On all other boundaries, no-slip BCs are applied.

The LBM parameters are identical for both showcases.
The fluid has a density of \SI{e3}{\kilogram\per\meter^3} and a kinematic viscosity of \SI{e-6}{\meter^2\per\second}.
TRT with $\Lambda = \Lambda_\text{mid}$ is chosen.
In order to sufficiently resolve the particles, the lattice spacing is ${\delta x=\SI{10}{\micro\meter}}$.
For the relaxation time ${\tau = 1.7}$, this results in a time increment of $\delta t = \SI{4e-5}{\second}$.
The density of the particles is only 14\% higher than the fluid density, and thus gravity is neglected.
Moreover, the lubrication correction according to \Sect{SubSec:LubrTheory} is applied for showcase~\ref{CP_SeparShowcase}.

For showcase~\ref{CP_AgglomShowcase}, the homogeneously charged spheres with \SI{8000}{\elementarycharge} (elementary charge \si{\elementarycharge}) have a radius of \SI{60}{\micro\meter}.
The particles are initially arranged as four sets of $6^3$ equi\-distantly placed spheres
with center-to-center distance \linebreak[2]\SI{120}{\micro\meter} in each dimension, as can be seen in \Fig{fig:VidSetupInit}.
The electric field is perpendicular to the flow, \ie{}, top and bottom plane impose the field via Dirichlet BCs for the potential.
In the simulation, the potential is set to ${\restr{\Phi}{\Gamma_B} = \SI{-100}{\volt}}$ at the (red) bottom plane and to
${\restr{\Phi}{\Gamma_T} = \SI{0}{\volt}}$ at the (blue) top plane.
All other boundaries are modelled as insulators, \ie{}, homogeneous Neumann BCs
$\restr{\frac{\partial \Phi}{\partial \vec{n}} }{\Gamma \backslash \left( \Gamma_B \cup \Gamma_T \right) } = 0$.\\
The parameters for this showcase, and the sets of spheres,
are also used for the scaling experiments of~\Sect{Sec:PerfMeas}.

For showcase~\ref{CP_SeparShowcase}, the homogeneously charged spheres with radius \SI{80}{\micro\meter} are inserted into the domain at random positions near the front wall.
Their charge is set to $\pm$\SI{40000}{\elementarycharge} with randomly generated sign.
The insertion frequency is chosen such that 4\% solid volume fraction is reached, under the assumption that the spheres move with the average fluid inflow velocity.
A homogeneous electric field is imposed in vertical direction via Dirichlet BCs at the channel front section.
The (blue) bottom plane is negatively charged with ${\restr{\Phi}{\Gamma_{B,f}} = \SI{-76.8}{\volt}}$, and the (red) top plane positively with ${\restr{\Phi}{\Gamma_{T,f}} = \SI{76.8}{\volt}}$.
To reduce particle agglomeration at the charged planes, homogeneous Neumann BCs are modelled at the top and bottom plane after the channel is split in two parts. 
These BCs are again applied to all other boundaries.\\

In~\Fig{fig:VidSetup}, the initial setup is shown for the agglomeration simulation, together with a later state.
The Poiseuille flow profile gets visible in~\Fig{fig:VidSetupImg167} from the positions of the equally charged particles
that are attracted by the bottom surface while they repel each other.

In \Fig{fig:SeparVidSetup}, simulation results are shown for the separation showcase.
\Fig{fig:SeparVid250PathlParticles} shows the particle distribution after \num{75300} time steps.
The particles move with the flow in longitudinal direction and are separated by the electric field.
While the positively charged (red) particles move towards the bottom plane and the lower half of the separated channel,
the negatively charged (blue) particles move in direction of the upper half.\\
In order to illustrate the particle movement, pathlines of each particle are depicted in the color indicating their charge.
To visualise the trajectories of all particles, and to evaluate how well particles are separated,
all pathlines at the end of the simulation are shown in~\Fig{fig:SeparVidAllPathlines}.
Here, one can see that almost all particles move to the correct subchannel intended for their charge, independent of their initial positions.
The particles that are not sorted correctly, \ie{} positively charges spheres entering the upper half of the separated channel and 
negatively charged particles entering the lower half, were hindered by other particles moving in opposite vertical direction.

The simulations were performed on 144 processes of the LiMa cluster of the computing center RRZE in Erlangen, Germany.
Showcase~\ref{CP_AgglomShowcase} comprises \num{128000} time steps that are carried out in \SI{10.8}{\hour} runtime.
For showcase~\ref{CP_SeparShowcase}, \num{210000} time steps are executed within \SI{15.7}{\hour}.
This performance, however, is degraded by writing visualisation data to disc,
and by additional global communication required for the insertion and deletion of particles in showcase~\ref{CP_SeparShowcase}.
Thus, the computational performance is further evaluted in \Sect{Sec:PerfMeas}.

\section{Performance Evaluation} \label{Sec:PerfMeas}

\subsection{Simulation Environment and Setup \label{SubSec:PerfMeasSetup}}
The computational performance is evaluated on the SuperMUC supercomputer at LRZ.
This cluster comprises 18 thin islands with 512 compute nodes each, connected by a high speed InfiniBand FDR10 interconnect.
Each node contains two Intel Xeon E5-2680 ``Sandy Bridge-EP`` octa-core processors that are running at 2.5 GHz, and 32 GB DDR3 RAM.
The code is built with the Intel 13.1 compiler,
IBM MPI 1.3, and Boost 1.49.
We initially show that the maximal overall performance is achieved with 16 MPI processes per node.
This number of processes is used also in the later experiments that employ several nodes.

\begin{figure}[ht]
  \centering
  \includegraphics[width=0.48\textwidth,page=1]{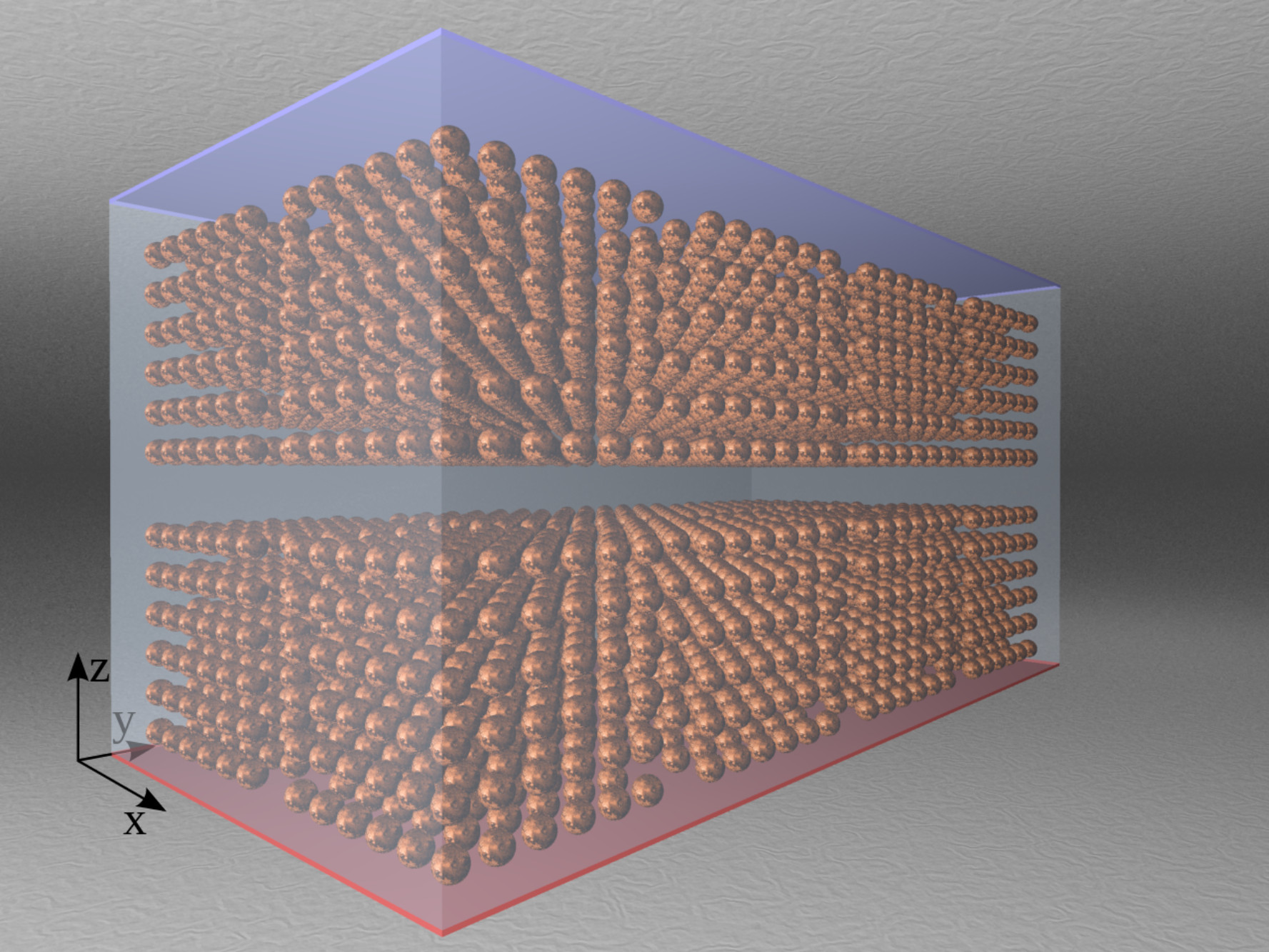}
    \caption{Charged particle simulation setup on a single node for weak scaling experiments.}
  \label{fig:WeakScalSetup}
\end{figure}

For the scaling measurements, the parameters described in~\Sect{sec:ModelProblem} are used.
A block of size $128^3$ cells 
with one set of spheres of radius $R=6 \delta x$ is assigned to each process.
For the single-node performance evaluation, the problem is extended once in x-direction and then only in y-direction until all 16 cores are occupied.
For the weak scaling experiments, the problem size per node is kept constant.
The setup on a fully occupied node with $256 \times 512 \times 256$ lattice cells and 3456 particles is shown in~\Fig{fig:WeakScalSetup}.
The weak scaling problem size is extended successively in all three dimensions, retaining maximal extension in longitudinal direction (to keep the aspect ratio of the domain comparable).
The maximum problem size on 2048 nodes comprises  $68.7 \cdot 10^9$ lattice cells and $7.08 \cdot 10^{6}$ particles.

About 9.4\% of the domain is covered by moving obstacle cells. These cells are not updated by the LBM.
Nevertheless, they are used for the potential computation. We define the measure
\emph{million lattice updates per second} (MLUPS) as the number of cells the Poisson problem
can be solved for within one second.
This metric indicates the performance to obtain the solution up to a given accuracy.
In the case of the LBM we refer to \emph{million fluid lattice updates per second} (MFLUPS)~\cite{wellein2006single} considering only fluid cells.

\begin{figure}[h!]
  \centering
  \includegraphics[width=0.48\textwidth,page=1]{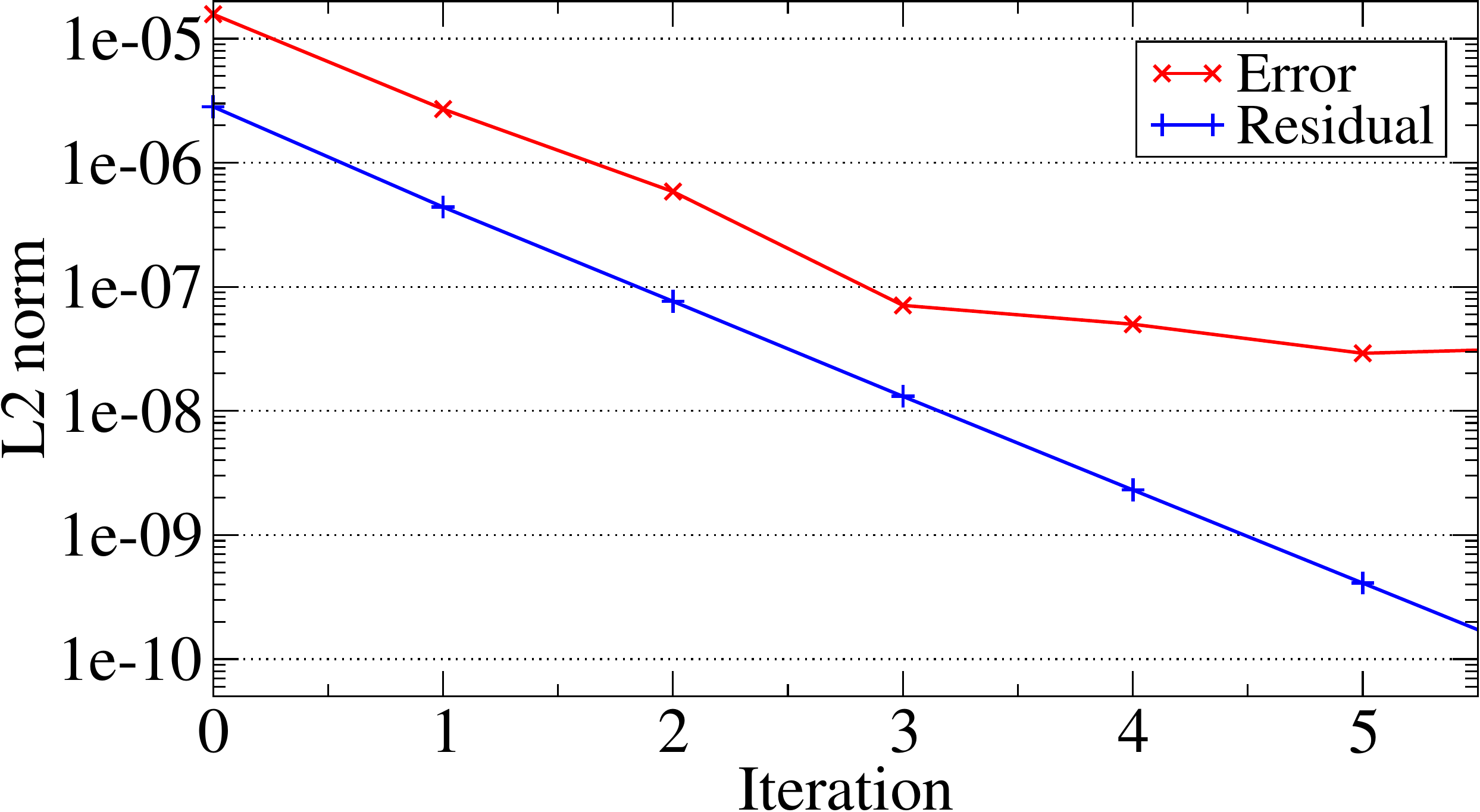}
    \caption{Error and residual after several V(2,2)-cycles}
  \label{Fig:ErrVsResid}
\end{figure}

Appropriate to the sphere radius, we use a subsampling factor of two for both, setting the RHS of Poisson's equation and the electrostatic force computation.
To find a suitable residual threshold for the MG solver, the $L_2$ norms of the residual and error are compared for several V(2,2)-cycles in \Fig{Fig:ErrVsResid}.
The error is strongly reduced up to a residual norm $1.5\cdot10^{-8}$.
Once the residual norm is smaller than $10^{-9}$, it is hardly reduced at all.
In the scaling experiments, V(3,3)-cycles are performed on seven levels, 
with $2^3$ cells per block on the coarsest level.
At the first time step, an initial solution is computed, requiring five V(3,3)-cycles.
Later steps only update the previous solution, requiring one V(3,3)-cycle.
The corresponding average $L_2$ residual norm over all time steps is about ${3\cdot10^{-9}}$.
This norm is monitored and never exceeds ${1.5\cdot10^{-8}}$.

\subsection{Single-Node Performance}
Before evaluating the scaling behaviour on several islands, we examine the single-node weak scaling performance.
An increasing number of processes is allocated to the processors in a round-robin fashion,
and 240 time steps are performed on the problem described in~\Sect{SubSec:PerfMeasSetup} ($128^3$ cells per block).
The problem on the coarsest grid with up to 128 cells (16 blocks, 7 MG levels) is solved with 6 CG iterations.
This number of iterations corresponds to the diameter of the maximal problem size on the coarsest grid.

\begin{figure}[htp]
  \centering
    \includegraphics[width=0.48\textwidth]{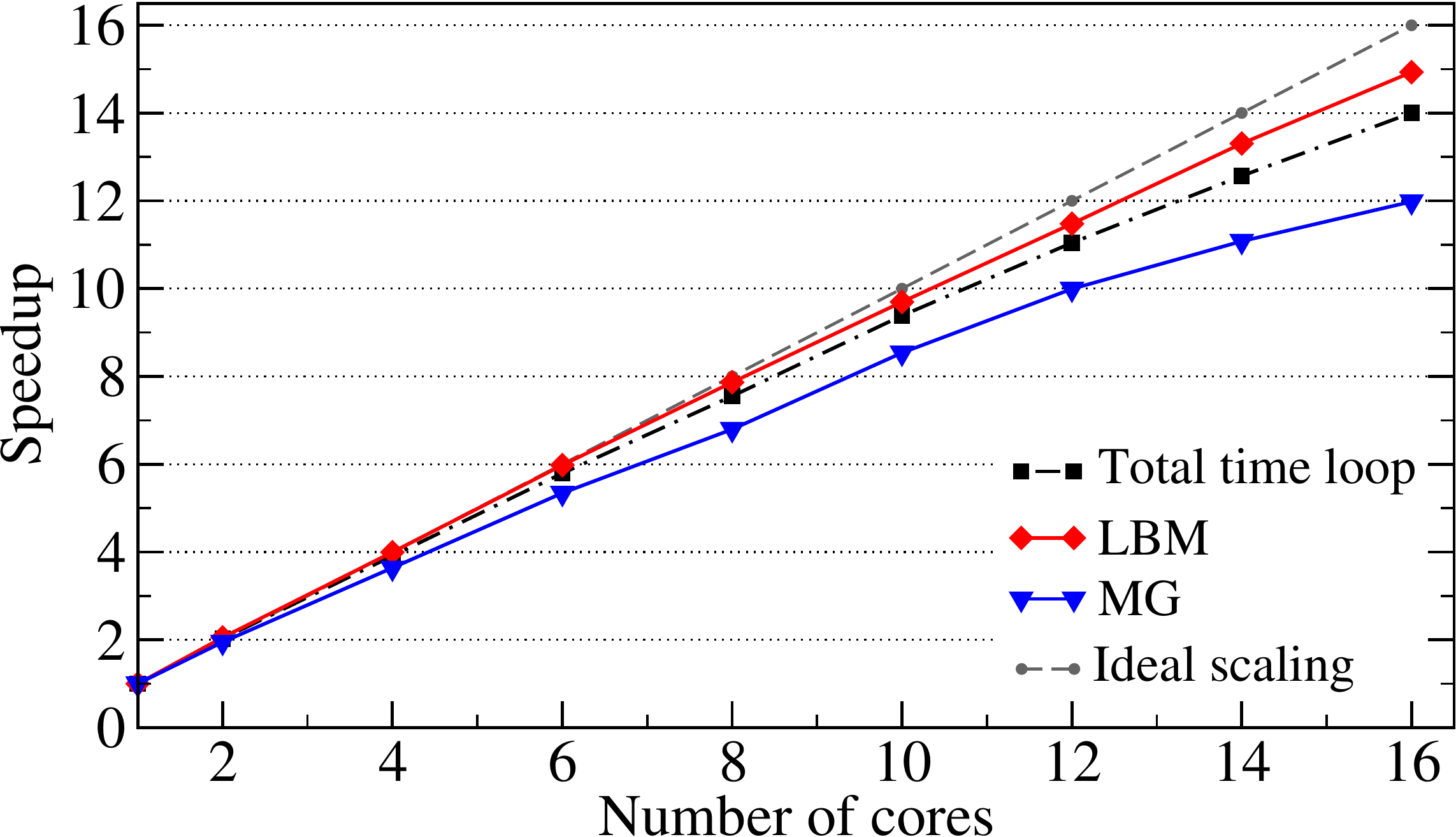}
    \caption{Single-node weak scaling performance of MG and LBM sweep for 240 time steps on SuperMUC.}
  \label{fig:SingleNodeWeakScalSMC}
\end{figure}

The total runtime increases from \SI{258}{\second} on one core to \SI{294}{\second} on 16 cores. 
This is equivalent to 88\% parallel efficiency of the overall algorithm.
\Fig{fig:SingleNodeWeakScalSMC} shows the speed\-up of the whole algorithm and its main parts for up to 16 MPI processes.
The LBM scales almost ideally and achieves 93\% parallel efficiency. Its performance is restricted only by computation-related data transfers, with negligible communication between the processes.
MG performance is additionally reduced by up to 10\% relative share of intra-node communication on the MG runtime,
resulting in 75\% parallel efficiency on a full node.

The single-node performance of the LBM sweep is 51.4 MFLUPS, and of the MG sweep 91.8 MLUPS.
These values will serve as base performance in the following section.
We employ 16 processes per node because this leads to the highest overall performance.

Considering the complicated physics simulated in the present paper,
both the LBM and the MG performance compare favourably to the results 
in~\cite{godenschwager2013framework} and~\cite{gmeiner2014peta}, respectively.\\
About $2 \times 78$ M(F)LUPS are achieved in~\cite{godenschwager2013framework}
for a pure LBM kernel on a SuperMUC compute node, utilizing SIMD and running at 2.7 GHz.
While the pure LBM can fully exploit the main memory,
the multiphysics simulations have to fit data structures for various fields into memory. 
The resulting reduced block size leads to a slight deterioration of the performance (\cf{}~\cite{Goetz:2010:ParComp}).
Moreover, the moving obstacles degrade the performance further as they cause a non-consecutive structure of fluid cells,
leading to an irregular memory access pattern.\\
The highly optimized SuperMUC single-node performance reported for the finite element MG solver in~\cite{gmeiner2014peta} corresponds to 170 MLUPS.
This performance is measured for V(3,3)-cycles and is applied to a Poisson problem with Dirichlet BCs.

\subsection{Weak Scaling}
We perform weak scaling experiments for up to 2048 nodes on SuperMUC for 240 time steps.
The problem size is successively doubled in all dimensions, as shown in~\Tab{Tab:WeakScalProbSizeCG}.
\begin{table}[hb]
   \centering
   \caption{Number of required CG coarse grid iterations for different problem sizes. \label{Tab:WeakScalProbSizeCG}}
   \resizebox{0.49\textwidth}{!}{%
   \begin{tabular}[h]{rcc|rcc}
      \toprule
      \#n.     & size & \#iter.          & \#n. & size & \#iter. \\
      \hline
      1     & $2  \times 2    \times 4$  &  6   &   64  & $8  \times 8    \times 16$ & 26   \\
      2     & $2  \times 2    \times 8$  & 10   &  128  & $8  \times 8    \times 32$ & 52   \\
      4     & $4  \times 2    \times 8$  & 12   &  256  & $16 \times 8    \times 32$ & 54   \\
      8     & $4  \times 4    \times 8$  & 12   &  512  & $16 \times 16   \times 32$ & 54   \\
      16    & $4  \times 4    \times 16$ & 24   & 1024  & $32 \times 16   \times 64$ & 114  \\
      32    & $8  \times 4    \times 16$ & 26   & 2048  & $32 \times 32   \times 64$ & 116  \\
      \bottomrule
   \end{tabular}
   }
\end{table}
The number of CG iterations required to solve the coarsest grid problem is depicted for different problem sizes.
When doubling the domain in all three dimensions, the number of CG iterations approximately doubles. 
This corresponds to the expected behaviour that the required number of iterations scales with the diameter of the problem size~\cite{2011:Gmeiner}, according to the growth in the condition number~\cite{Shewchuk94}.
However, when doubling the problem size, CG iterations sometimes stay constant or have to be increased.
This results from different shares of Neumann and Dirichlet BCs on the boundary.
Whenever the relative proportion of Neumann BCs increases, convergence deteriorates and more CG iterations are necessary. 

\begin{figure}[hb]
   \centering
   \includegraphics{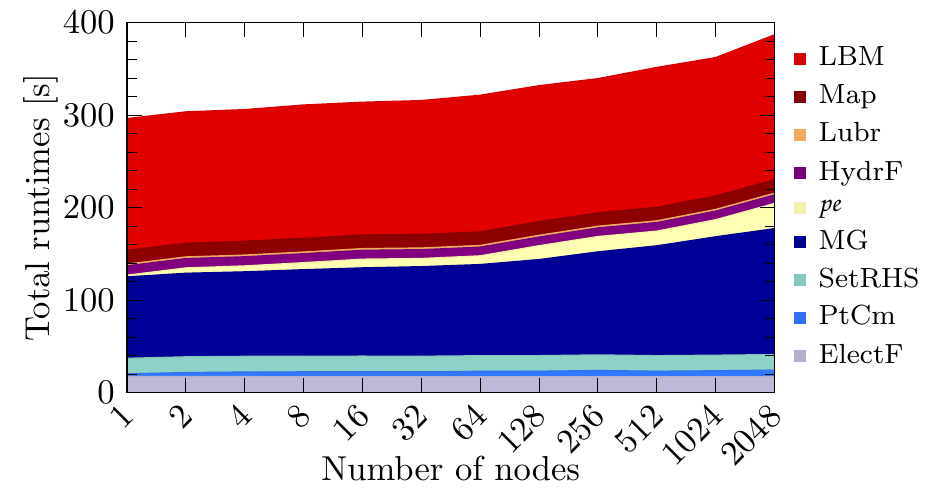}
    \caption{Runtimes of charged particle algorithm sweeps for 240 time steps for an increasing number of nodes.}
   \label{fig:WeakScalSharesOverallAlg_240TS_SMC}
\end{figure}

The runtimes of all parts of the algorithm are shown in \Fig{fig:WeakScalSharesOverallAlg_240TS_SMC} for different problem sizes,
indicating their shares on the total runtime.
This diagram is based on the 
maximal (for MG, LBM, \pe{}) or 
average (others) runtimes of the different sweeps among all processes.
The upper part of the diagram shows the cost of fluid-simulation related sweeps, such as LBM, moving obstacle mapping (Map), hydrodynamic force computation (HydrF), and lubrication correction (Lubr) sweep.
In the middle, the cost of the \pe{} sweep is shown.
Below, the costs of sweeps related to the electric potential computation are displayed.
These include MG, setting the RHS of Poisson's equation (SetRHS), communication of the electric potential before the gradient computation (PtCm), and the sweep computing the electrostatic force (ElectF).

For a more precise evaluation, the exact figures are shown in \Tab{Tab:WeakScalTimeShare} for one node and 2048 nodes.
The total runtime (Whl) is less than the sum of the individual sweeps, since different sweeps are slow on different processes.
\begin{table}[h!t]
  \centering
  \caption{Time of the whole algorithm and its sweeps for 240 time steps on a single node and on 2048 nodes. \label{Tab:WeakScalTimeShare} }
\resizebox{0.49\textwidth}{!}{%
  \begin{tabular}[h]{r|c|ccccc}
   \toprule
 & Whl    & LBM   & MG & \pe{} & PtCm & Oth \\
\#n.      & t$\left[ s \right]$ & t$\left[ s \right]$ ($\left[ \% \right]$) & t$\left[ s \right]$ ($\left[ \% \right]$) & t$\left[ s \right]$ & t$\left[ s \right]$ & t$\left[ s \right]$ \\
 \hline
     1       & 294    & 143 (48) & 88 (30)   & 2   &  3  & 61 \\
     2048 & 353    & 157 (41) & 136 (35) & 27 &  7  & 60 \\
    \bottomrule
  \end{tabular}}
\end{table}
Sweeps whose runtimes depend on the problem size---mainly due to increasing MPI communication---are LBM, MG, \pe{}, and PtCm.
Overall, LBM and MG take up more than 75\% of the total time, w.r.t. the runtimes of the individual sweeps.

The sweeps that scale perfectly---HydrF, LubrC, Map, SetRHS, and ElectF---are summarized as `Oth`.
For longer simulation times the particles attracted by the bottom wall are no longer evenly distributed, possibly causing load imbalances.
However, they hardly affect the overall performance.
For the simulation of showcase~\ref{CP_SeparShowcase}, the relative share of the lubrication correction is below 0.1\%, and each other sweep of `Oth` is well below 4\% of the total runtime.

\begin{figure}[htp]
  \centering
    \includegraphics[trim = 0mm 1mm 0mm 1mm, clip,width=0.49\textwidth]{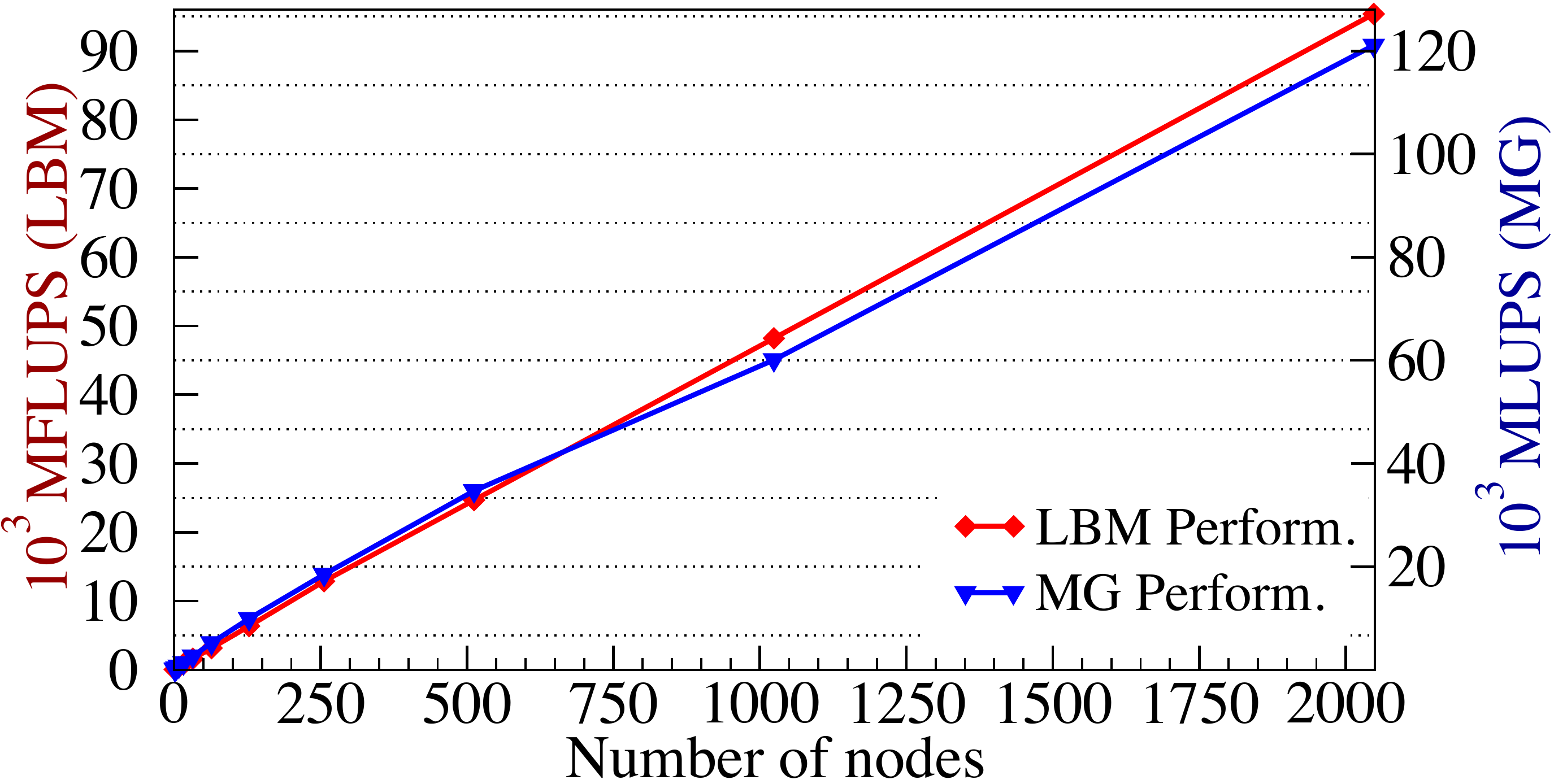}
        \caption{Weak scaling performance of MG and LBM sweep for 240 time steps.}
  \label{fig:WeakScalMLUPs240TS_SMC}
\end{figure}

Overall, the coupled multiphysics algorithm achieves 83\% parallel efficiency on 2048 nodes.
Since most time is spent to execute LBM and MG, we will now turn to analyse them in more detail.
\Fig{fig:WeakScalMLUPs240TS_SMC} displays the parallel performance for different numbers of nodes.
On 2048 nodes, MG executes 121,083 MLUPS, corresponding to a parallel efficiency of 64\%.
The LBM performs 95,372 M\-FLUPS, with 91\% parallel efficiency.

\begin{figure}[h!]
   \centering
   \includegraphics{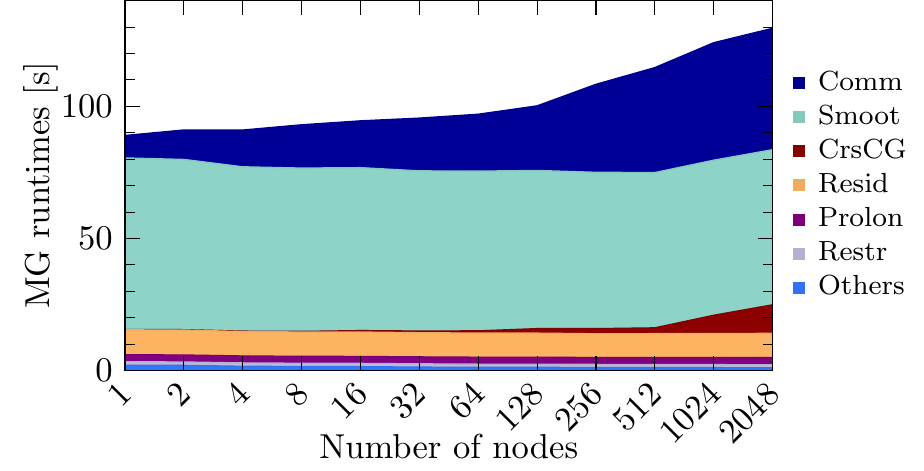}
    \caption{Average runtime of different MG sweeps for 240 time steps for an increasing number of nodes.}
  \label{fig:WeakScalSharesMG_240TS_SMC}
\end{figure}

Finally, the average execution times of different MG sub-sweeps are presented in \Fig{fig:WeakScalSharesMG_240TS_SMC}.
The parallel performance is degraded in particular by communication (Comm), but also by the CG coarse grid solver (CrsCG) due to the increasing number of required iterations.
Clearly, the CG-based coarsest grid solver is a non-scalable part of the solver that will ultimately dominate the performance.
However, for computations up to the size presented in this paper, the overhead is still in an acceptable range.

All other sweeps show ideal scaling. 
Among these sweeps, most time is spent for smoothing (Smoot)
and residual computation (Resid).
All other parts require only a small portion of the overall MG time.
These are the sweeps for
prolongation (Prolon, \SI{2.7}{\second}), restriction (Restr, \SI{1.2}{\second}), and the sweeps summarised as `Other`,
\ie{}, those for
adapting the RHS to the BCs (about \SI{1}{\second}) and
setting the solution on the following coarser grid with zero (\SI{0.2}{\second}).
The sweeps for checking the termination criterion
and adapting the stencils to the BCs at the beginning of the simulation
are negligible ($<\SI{0.01}{\second}$).
%

%% file: conclusion.tex
\section{Conclusion\label{sec:Conclusion}}
In this article we demonstrate the potential of coupled multiphysics simulations
using the LBM, a physics engine, and a scalar potential solver.
These components are integrated into the parallel software framework  \Walberla{}
to simulate
fully resolved charged particles in microfluidic flows.
We describe the software design,
validate the physical correctness for different model parameters,
and present the excellent computational performance and scalability of the algorithms.
The results show that the multiphysics algorithm allows the
physically correct simulation of several millions of interacting charged particles,
as required for advanced realistic multiphysics simulations.

For the presented
simulations, \Walberla{} is extended by
a parallel cell-centered MG method to compute the electrostatic potential including an
efficient handling of the boundary conditions.
The code is carefully optimised by exploiting that the stencils on the finest level are (almost) constant
and thus the main memory access can be reduced,
while retaining the flexibility of Galerkin coarsening.

The fluid simulation and the fluid-particle coupling
are validated for low Reynolds number flows 
for spheres arranged in a periodic array.
The largest relative error for the drag is shown to be below 2.2\% for 
various sphere resolutions, lattice viscosities, and values of the `magic' TRT parameter $\Lambda$.
For commonly used parameters and sphere resolutions, the relative error is around 1\% or even below.
We further observe that the value of the parameter $\Lambda$ that is optimised for porous media yields slightly better results
than the alternative value which is chosen to set the boundary mid-way between two lattice sites.
For the higher examined lattice viscosity the error can be reduced further.
The error analysis is extended to an evaluation of volume mapping errors, and estimates of the drag that is corrected taking these errors into account. Furthermore, the effect and the correctness of the lubrication correction method
are shown for different sphere resolutions, sphere velocities, and lattice viscosities
at low Reynolds numbers.
The influence of these parameters on the lubrication correction is analysed carefully
in the situation of sphere-sphere and sphere-wall interaction.
This confirms that for the TRT model the lattice viscosity has only a minor impact.\\
To validate the interaction with electrostatic forces, the computed electric potential and the resulting force on a single charged sphere
are analysed for different sphere resolutions.
A subsampling of the sphere geometry is used to reduce
volume mapping errors and thus to improve the accuracy of the
electrostatic force computation,
while keeping the mesh conforming to the LBM and without increasing the
systems that must be solved.
The relative error in the computed electric potential decreases with increasing sphere resolutions and with more levels of subsampling,
however, the error in the electric potential
stagnates once the finite volume discretisation error is reached.
The largest relative error of the potential
is well below 2\% for the resolutions presented.
Subsampling is found to have more influence on the electrostatic force than on the electric potential. 
The subsampling for the electrostatic force computation is found to be necessary
only for sphere radii below nine lattice spacings to obtain errors that are of comparable magnitude to those arising from
the fluid-particle interaction.

The parallel performance of the multiphysics algorithm is examined on SuperMUC.
The overall algorithm has a parallel efficiency of 83\% on 32,768 cores, simulating more than seven million charged particles.
Most time of the algorithm is spent for the LBM and the MG solver.
For both, the single-node performance compares favourably to optimised reference implementations of pure LBM and finite element MG.
The MG scales with 64\% parallel efficiency, achieving more than $121\cdot 10^9$ cell updates per second to solve Poisson's equation with mixed boundary conditions.
Here, the reduced scalability results from an increasing number of coarsest grid solver iterations and increasing MPI communication, with many small messages on the coarse grids.
The LBM for fluid-particle interaction scales almost perfectly with 91\% parallel efficiency, achieving more than $95\cdot 10^9$ fluid cell updates per second.

The example scenarios and their animations demonstrate the qualitative correctness of the overall algorithm and hint at future applications.
The present algorithm can be applied to find design parameters, \eg{} for optimal separation efficiency of oppositely charged particles as in the separation scenario,
depending on fluid and particle properties.
Moreover, the algorithm allows future 
simulations of large numbers of charged particles in more complex geometries, such as electrostatic filters or pulmonary drug delivery, also for higher Reynolds numbers.\\
The MG solver allows to simulate multiphysics applications beyond those in this article, 
such as simulations comprising temperature fields described by Poisson's equation.
Also jumping dielectricity coefficients can be realised easily to simulate dielectrophoresis.
The solver module and BC treatment can be applied to more complicated elliptic partial differential equations,
such as the Poisson-Boltzmann equation.
This equation allows to simulate charged particles in electrolyte solutions, including double-layer effects 
that are of relevance for simulating lab-on-a-chip systems.